\documentclass[reprint,amsmath,amssymb,showkeys,longbibliography,aps,pra,floatfix,onecolumn,superscriptaddress]{revtex4-2}
\usepackage{graphicx}
\usepackage{dcolumn}
\usepackage{bm}
\usepackage[hidelinks]{hyperref}
\usepackage{general_latex}
\usepackage{tabu}
\usepackage{caption}
\usepackage{subcaption}
\usepackage[normalem]{ulem} 
\usepackage{setspace} 
\def\c{\mathrm{c}}

\def\s{\mathrm{s}}
\def\BdG{\mathrm{BdG}}
\def\zero{\mathrm{zero}}
\def\GE{\mathrm{ge}}
\def\col{\mathrm{col}}
\def\con{\mathrm{con}}
\def\dis{\mathrm{dis}}

\DeclareMathOperator*{\SumInt}{%
\mathchoice%
  {\ooalign{$\displaystyle\sum$\cr\hidewidth$\displaystyle\int$\hidewidth\cr}}
  {\ooalign{\raisebox{.14\height}{\scalebox{.7}{$\textstyle\sum$}}\cr\hidewidth$\textstyle\int$\hidewidth\cr}}
  {\ooalign{\raisebox{.2\height}{\scalebox{.6}{$\scriptstyle\sum$}}\cr$\scriptstyle\int$\cr}}
  {\ooalign{\raisebox{.2\height}{\scalebox{.6}{$\scriptstyle\sum$}}\cr$\scriptstyle\int$\cr}}
}

\begin{document}
\preprint{APS/123-QED}
\title{Quantum Backreaction effect in Optical Solitons}

\author{Sang-Shin Baak}
\affiliation{School of Physics and Astronomy, SUPA, University of St.~Andrews,\\
North Haugh, St. Andrews KY16 9SS, United Kingdom
}
\affiliation{Seoul National University, Department of Physics and Astronomy, Center for Theoretical Physics, Seoul 08826, Korea} 
\author{Friedrich K\"onig}
\affiliation{School of Physics and Astronomy, SUPA, University of St.~Andrews,\\
North Haugh, St. Andrews KY16 9SS, United Kingdom
}
\date{\today}
\begin{abstract}
   Optical solitons classically are stationary solutions of the nonlinear Schr\"odinger equation.
    We perform a quantum field theoretic treatment by quantising a linearised fluctuation field around the classical soliton solution which can be seen as providing a background spacetime for the field. The linearised fluctuation modifies the soliton background, which is often neglected, reminiscent of the \emph{nondepleted-pump} approximation. Going beyond this approximation and by using a number-conserving Bogoljubov approach, we find unstable modes that grow as the soliton propagates. Eventually, these unstable modes induce a considerable (backreaction) effect in the soliton. We calculate the backreaction in the classical field fully analytically in the leading second order. 
    The result is a quadratic local decrease of the soliton photon number in propagation due to the backreaction effect of the unstable mode. 
    Provided the initial pulse is close to the classical soliton solution, the unstable mode contributions always become dominant. 
    We also consider practical scenarios for observing this quantum-induced soliton distortion, in the spectral domain. The backreaction, which we expect to be present in bright and dark, discrete and continuous solitons and other nonlinear pulses plays an important role for future optical analogue gravity experiments, for soliton lasers, and optical communications.
\end{abstract}
\maketitle
\tableofcontents
\newpage
\section{Introduction}
\label{sec:Intro}
Advancement in physics is crucially dependent on the unification of gravity and quantum theory in a theory of quantum gravity.
Recently, a number of theoretical studies identified significant effects at the interplay between classical gravity and nonclassical quantum fields \cite{Hollands2015}. 
The coupling of a massive quantum field with a classical background field is described in quantum field theory in curved spacetime (QFTCS) \cite{Fulling1989,Kay2023}.
In this, the kinematics of the quantum field are determined by the background spacetime, and QFTCS has discovered many important phenomena such as Hawking radiation, cosmological particle creation, the dynamical Casimir Effect, and the Unruh effect \cite{Wald1994,Mukhanov2007,Ford1997,Parker2009,Birell1982}.
These phenomena depend on how the matter fields evolve on a given curved spacetime background.
Going beyond, general relativity describes how matter fields determine the dynamics of spacetime \cite{Wald1984,Hawking1973,MTW2017}.
Therefore, we also need to describe the impact of a quantum field onto a background spacetime.\\

In this setting, we first need to set the background, which evolves according to its own dynamics. Next, we plant an ``object'' (field) onto the background, which will interact with the background and modify its dynamical evolution. This modification is what we call the backreaction. Therefore, the backreaction is dependent on the choice of both the background and the object field. For example, in cosmology the object might be the inhomogeneous matter of our universe such as galaxies, dark matter, dark energy, etc. The backreaction then refers to the difference between the observed universe and the Freedmann-Lema\^itre-Robertson-Walker universe \cite{Schander2021}.
In another example, semi-classical gravity, the background is chosen to be the spacetime evolving with classical dynamics whose source includes the classical matter distribution and objects are the quantum fields residing in it. In this case, the expectation value of the energy-momentum tensor, the source term in the semi-classical Einstein equation, is regarded as a backreaction.
The investigation of the backreaction of quantum fields in gravity is interesting, as it can serve as a next-order approximation to quantum gravity with observable effects. \\

Although significant progress has been made on the theory, the observation of QFTCS phenomena or the backreaction in real gravity has been elusive because of the large scale or the weakness of their signature.  In his paper in 1981, Unruh predicted that Hawking radiation or the evaporation of black holes can be investigated in the laboratory in fluids using sound waves \cite{Unruh1981}.
His paradigmatic idea of a waterfall, a fluid accelerating beyond the speed of sound, representing an effective black hole spacetime signposted how to investigate QFTCS experimentally in what became known as the field of `analogue gravity'.
The key insight is that the Hawking effect is kinematic and therefore reproducible in non-gravitational systems.
The linear perturbations of many physical systems can be shown to reside on an effective spacetime named analogue spacetime.
Typical analogue spacetimes investigated experimentally are of black holes or cosmic expansion. 
Arguably, the most celebrated result to date is the observation of the Hawking effect in various systems in the lab \cite{Philbin2008,Silke2011,Steinhauer2016,Steinhauer2019}.
Going beyond the linarised description, there is a growing interest in the backreaction in analogue gravity systems \cite{Fischer2005,Fischer2007,Balbinot2005PRD,Balbinot2005PRL,Fagnocchi2006,Schuetzhold2008JPA,Schuetzhold2008PoS,Kurita2010,Tricella2020,Patrick2021,Baak2022,Braunstein2023}.
In an analogue spacetime, the quantum dynamics of the full system (containing the quantum- and background fields) is known via its Hamiltonian.
Hence, the backreaction can be extracted from the full quantum dynamics of the system. 
Although the background dynamics generally are different compared to gravity, the backreaction in analogue gravity allows us to investigate spacetime dynamics beyond the limited simulations on classical computers and establishes novel quantum-classical systems.\\

One of the most promising physical systems in analogue gravity is the fiber-optical system.
The propagation of electromagnetic waves in an optical fibre is well described by the nonlinear Schr\"odinger (NLS) equation. 
This equation features stable soliton solutions which have been used for investigating a 1+1-dimensional black hole spacetime before \cite{Philbin2008, Choudhary2012}.
The backreaction in this system, however, has not been systematically formulated to date, although  a few descriptions go beyond the linearisation approach \cite{Villari2018,Ward2023}.
To describe the backreaction, we first set the background and object for the system.
The soliton is our background and the linear quantum fluctuation on it is the object.
The dynamics of the linear order does not include the backreaction and to find backreaction effects, the dynamics need to be treated at least in the second order.
A well-known technique in atomic physics to treat the system beyond linear perturbation order is the Number-Conserving Bogoliubov expansion (NCBE). 
NCBE is the field expansion scheme beyond linearisation which uses the number-conservation as a guiding principle \cite{Girardeau1959,Gardiner1997,Girardeau1998,Castin1998,Lieb2000,Gardiner2007,Billam2012,Billam2013}.\\

In this paper, we introduce a mean field correction which is second-order in our expansion parameter. We establish that the correction is the leading order backreaction effect on the soliton. It is consistent with an expansion we construct of the quantum field into a complete set of modes, including discrete modes. The relation to the standard linear soliton perturbation theory is presented.
While the linear order does not conserve the photon number, the field correction restores number-conservation in our working order and at the same time avoids rather involved coupled equations of motion. 
A typical problem of the full NCBE is that it is hard to get to a practical solution, because it results in a complicated set of coupled equations.
By our approach we obtain the general analytic solution to the field correction.
We show that as the pulse propagates, the field correction approaches a characteristic configuration independent of the initial configuration.\\

In Sec.~\ref{sec:NCExpansion}, we review the linearisation approach and its limitations and introduce the number-conserving expansion to the quantum soliton field. We proceed to recall properties of the pseudo-Hermitian Bogoliubov-de Gennes Hamiltonian and derive a complete mode expansion in
Sec.~\ref{sec:Bogoliubov}. Here we also calculate the photon number density.
Sec.~\ref{sec:Classical} derives analytical expressions for the classical field correction. 
Finally, in Sec.~\ref{sec:Physical}, we work out which measurable physical quantities can be extracted from the analysis. In particular, the behaviour of the total photon number  and the spectral density for long propagation distances is discussed, before concluding.

\section{Linearisation and the number-conserving expansion of the soliton field}
\label{sec:NCExpansion}
The quantum dynamics of optical solitons is described by the quantized NLS (QNLS) equation.
This equation is integrable and the exact solution can be obtained following a Bethe ansatz \cite{Lai2}.
It was shown that for large photon numbers, the soliton classical NLS solution acts as the mean field in leading order \cite{Lai1}. 
By quantizing a perturbation of this mean field, the interesting interplay between the classical and quantized NLS equations can be studied. 
In this approach, the object is the linearized quantum fluctuation of the field.
The transformation of the linear fluctuation through the interaction with the soliton mean field has been studied theoretically and experimentally and many interesting phenomena have been established such as soliton squeezing and entanglement \cite{Carter1987, Spalter1998PRL, Schmidt1998, Silberhorn2001, Korolkova2002, Carter1987, Korolkova2001, Haus1990, Imoto1985, Werner1997, Werner1996}
This approach, however, is inadequate for describing the backreaction, because the background evolution is not affected by the quantum fluctuation. It is reminiscent of the \emph{non-depleted pump} approximation in parametric down-conversion \cite{Adami2014,Couteau2018}\\

To describe the backreaction effect, we make use of the number-conserving approach. 
In this approach, we go beyond the linear expansion order by introducing a quadratic order correction to the classical field to recover photon number conservation in our working order. The classical field correction also is the backreaction.
In what follows, we first briefly review the QNLS equation and the linearisation approach. Starting with the classical Lagrangian for the NLS equation, we canonically quantise the field and derive the QNLS equation. We identify the equation of motion for both fields, the soliton $E_\s$ and the linear quantum fluctuation $\Delta E$.
We identify the expansion parameter and expand to quadratic order. Then we obtain the classical field correction $E_\c$ and derive an evolution equation for this field.

\subsection{Linearisation around the soliton solution to the quantum nonlinear Schr\"odinger equation}
\label{sec:ReviewLinear}
The classical electric field in a nonlinear optical fiber satisfies the nonlinear Schr\"odinger equation \cite{Agrawal2019}.
The Lagrangian for the NLS equation is \cite{Wright1991}
\begin{equation}\label{eq:NLSElagrangian}
    \mathcal{L} = \frac{i}{2}(E^*\partial_tE-\partial_tE^*E)-\frac{1}{2}|\partial_xE|^2+\frac{1}{2}|E|^4,
\end{equation}
where $E$ describes the slowly varying envelope field amplitude and we employ soliton units \cite{Agrawal2019}.
The fundamental soliton solution is of the form
\begin{equation}
    E_\s(t,x) = \sech x\,\mathrm{e}^{i t/2}.
\end{equation}
The solution famously has a stationary field envelope as it propagates along the fiber.

In order to quantise the equation of motion, we follow the canonical quantisation procedure.
From the Lagrangian \eqref{eq:NLSElagrangian}, we calculate the conjugate momenta of the electric field,
\begin{equation}
    \Pi_E = iE^*.
\end{equation}
The QNLS equation is obtained by promoting the electric field to a quantum field.
Hence, we impose the equal-time commutation relations
\begin{align}
    \frac{1}{i}[\hat{E}(t,x),\hat{\Pi}_E(t,x')]=[\hat{E}(t,x),\hat{E}^\dagger(t,x')]&=\delta(x-x'),\label{eq:EFieldCommutationRelation1}\\
    [\hat{E}(t,x),\hat{E}(t,x')]=[\hat{E}^\dagger(t,x),\hat{E}^\dagger(t,x')]&=0\label{eq:EFieldCommutationRelation2}.
\end{align}
The corresponding Hamiltonian for the QNLS equation is \cite{Wright1991}
\begin{equation}
    \hat{H} = \frac{1}{2}\int \ud x [\partial_x\hat{E}^\dagger\partial_x\hat{E} - \hat{E}^\dagger \hat{E}^\dagger\hat{E}\hat{E}],\label{eq:QNLSHamiltonian}
\end{equation}
where normal ordering is adopted and $\hbar=1$ for simplicity.
Finally, the equation of motion of the quantized envelope $\hat{E}$ is
\begin{equation}
    \left(i\partial_t+\frac{\partial_x^2}{2}+\hat{E}^\dagger\hat{E}\right)\hat{E}=0.\label{eq:QNLSE}    
\end{equation}
Analogous to the NLS equation, Eq. \eqref{eq:QNLSE} is called the quantum NLS (QNLS) equation.
The exact energy eigenvalues and eigenfunctions for this equation can be obtained following a Bethe ansatz.
The exact solutions, however, obtained with this ansatz do not easily relate to the fundamental soliton solution under consideration here.
It is known that the quantum dynamics of the NLS soliton can be described by the Hartree approximation provided the photon number is large \cite{Lai1,Lai2}.
This routinely used technique is a linearisation around the soliton and facilitates our investigation of the  the QNLS equation.

We now briefly introduce this linearisation approach.
The procedure starts with writing the quantized envelope field $\hat{E}$ as the sum of a fundamental soliton $E_\s$ and a small quantum fluctuation $\Delta \hat{E}$:
\begin{equation}\label{eq:Linearisation}
    \hat{E} \stackrel{!}{=} E_\s+\epsilon\Delta \hat{E}.
\end{equation}
where $\stackrel{!}{=}$ indicates that the expansion is only valid in the linearisation approach. $\epsilon$ is a small number for specifying perturbation. It will be clarified later.
Because the classical field commutes, the commutation relations \eqref{eq:EFieldCommutationRelation1} and \eqref{eq:EFieldCommutationRelation2} become
\begin{equation}
    [\Delta\hat{E}(t,x),\Delta\hat{E}^\dagger(t,x')]=\delta(x-x'),\qquad [\Delta\hat{E}(t,x),\Delta\hat{E}(t,x')] = 0.
\end{equation}
Next we recall that the soliton is the solution of the NLS equation, i.e.
\begin{equation}\label{eq:NLSE}
    \left(i\partial_t+\frac{\partial_x^2}{2}+|E_\s|^2\right)E_\s=0,
\end{equation}
which forms our zeroth order description. This equation does not contain a $\Delta\hat{E}$ contribution, i.e. the background soliton is unaffected by the quantum fluctuation.
Substituting the ansatz \eqref{eq:Linearisation} into the QNLS equation \eqref{eq:QNLSE} and using Eq.~\eqref{eq:NLSE}, the resulting equation is at least linear order in $\Delta \hat{E}$.
Provided $|\Delta E|^2\ll |E_\s|^2$, we may neglect higher orders and arrive at the so-called Bogoliubov-de Gennes (BdG) equation
\begin{align}
\left(i\partial_t+\frac{\partial_x^2}{2}+2|{E}_\s|^2\right)\Delta\hat{E}+ {E}_\s^2\Delta\hat{E}^\dagger=0.\label{eq:QNLSlinearise}
\end{align}
The equation with the linear approximation above describe the time evolution of the linear quantum fluctuation.
The linearisation approach has been successfully used for the quantum noise evolution or soliton squeezing \cite{Haus1990,Lai1993,Margalit1998,Haus2000}, but it
does not describe effects beyond the linear order \cite{Kaup1990,Kaup1991}.
The backreaction in analogue gravity, in which we are interested, is a quantity beyond linear order; and so we need a theory going beyond linear order.
The framework we use for this purpose is the number-conserving expansion, which we explain next.

\subsection{Beyond linearisation: number-conserving expansion}
\label{sec:BeyondLinear}
Before detailing the number-conserving expansion, we clarify the importance of photon-number conservation in the QNLS system.
The QNLS Hamiltonian \eqref{eq:QNLSHamiltonian} has a global $U(1)$-symmetry, leading to the following continuity equation for the QNLS field $\hat{E}$,
\begin{equation}
    \partial_t \rho +\partial_x J=0,
\end{equation}
where $\hat{\rho}$ corresponds to the photon-number density and $\hat{J}$ to the photon current:
\begin{align}
    \rho &=\langle\hat{\rho}\rangle = \langle\hat{E}^\dagger\hat{E}\rangle,\label{eq:densityoperator}\\
    J &=\langle\hat{J}\rangle =\frac{1}{2i}\langle\hat{E}^\dagger\partial_x\hat{E}-\partial_x\hat{E}^\dagger\hat{E}\rangle.\label{eq:currentoperator}
\end{align}
where $\langle \cdot \rangle$ is the quantum expectation value.
Consequently, the total photon number is conserved in time.
In addition, it can be used as an expansion parameter in the linearisation approach as follows.
Let us recall that $\int |E_s|^2 \ud x=2 =\mathcal{O}(1)$. Thus, all terms in the NLS equation \eqref{eq:NLSE} are of order one.
Moreover, all the terms in the BdG equation \eqref{eq:QNLSlinearise} are of the same order as $\Delta\hat{E}$.
Also recall that if the photon number is large, the exact solution goes to the fundamental soliton solution. 
Therefore, the natural choice of ratio between classical soliton field and its fluctuation is 
\begin{equation}\label{eq:ExpansionParameter}
    \frac{\langle\Delta \hat{E}^\dagger\Delta \hat{E}\rangle}{\langle\hat{E}^\dagger\hat{E}\rangle} \sim \frac{1}{N} =: \epsilon^2\ll1
\end{equation}
where $N$ is the total photon number of the system.
In other words, the inverse square root of the total photon number is the expansion parameter $\epsilon$ in the linearisation approach.
Therefore, it is important to maintain the number-conservation in the working order to have a well-defined expansion of the field \cite{Castin1998,Ralf2006}.

A number-conserving expansion is a popular way to obtain a field expansion beyond linearisation.
The number conservation can be maintained in each order by inclusion of an additional annihilation operator to keep the $U(1)$-symmetry in the expansion \cite{Girardeau1959,Girardeau1998,Gardiner1997,Ralf2006,Gardiner2007,Billam2012,Billam2013,Castin1998,Fischer2005}.
This often leads to complicated, coupled equations, which are hard to solve.
To avoid these complications, we modify the approach by limiting the number-conservation to our working order ($\sim\epsilon^2$).

In our approach, we want to obtain the classical field correction describing the backreaction effect.
Recall the classical electric field is regarded as the quantum expectation value of the quantum electric field \cite{Leonhardt2010}.
In the linearisation approach, this field is the classical soliton $E_s$. In the number-conserving approach we additionally introduce a 2nd order correction term $E_c$, a quantum correction to the classical field, which recovers number conservation.
It follows, that the field expansion may not lie around the soliton field $E_\s$, but needs a small correction. 
We use the conventional linearisation approach, breaking the $U(1)$-symmetry, and apply a correction term to the classical field which recovers the number-conservation at the desired order ($\sim\epsilon^2$) \cite{Baak2022}:
\begin{equation}\label{eq:electricfieldexpectation}
    \langle\hat{E}\rangle = E_\s+\epsilon^2E_\c.
\end{equation}
The number-conserving expansion for the quantized envelope field is therefore 
\begin{equation}
    \hat{E} = E_\s+ \epsilon\Delta \hat{E}+\epsilon^2E_\c
    \label{eq:orderexpansion}
\end{equation}
The equation of motion can be obtained by expanding the QNLS equation and collecting terms in orders of $\epsilon$, respectively.
Since the correction term is of order $\epsilon^2$, it does not affect the leading-order and linear-order dynamics.
In other words, the leading-order expansion gives the NLS equation \eqref{eq:NLSE}, and therefore the soliton is still the leading-order solution of the classical field.
The linear fluctuation $\Delta\hat{E}$ satisfies the BdG equation \eqref{eq:BdG}.
The quadratic order term, $E_\c$, which corresponds to the correction of the classical electric field, satisfies an inhomogeneous equation with $\Delta \hat{E}$-dependent source terms. It has a  BdG-like homogeneous part:
\begin{align}
\left(i\partial_t+\frac{\partial_x^2}{2}+2|E_\s|^2\right)E_\c+ E_\s^2E_\c^{*}=-2\langle\Delta\hat{E}^\dagger\Delta\hat{E}\rangle E_\s-\langle\Delta\hat{E}^2\rangle E_\s^*.\label{eq:GPcorr}
\end{align} 
Note that the right hand side of the equation contains the expectation value of the quantum fluctuation in the background soliton.
Therefore, $E_\c$ is the leading order backreaction effect and we call it the quantum backreaction (QBR) equation.

\section{The Bogoliubov-de Gennes equation}
\label{sec:Bogoliubov}
In the previous section we derived a set of equations for describing the backreaction for a classical soliton background. In this section, we solve the Bogoliubov-de Gennes equation \eqref{eq:QNLSlinearise}  
such that we expand the field in its normal modes. We also calculate vacuum expectation values relevant in section~\ref{sec:Classical}.

We start by considering the mathematical properties of the BdG Hamiltonian, which produces the Bogoliubov-de Gennes equation \eqref{eq:QNLSlinearise}, because it allows us to understand the structure of eigenvalues and eigenvectors of the BdG solutions.
In Sec.\ref{sec:MathBdG} below, we summarize its nontrivial structure, which lies beyond Hermitian quantum mechanics.
Then, we derive the mode structure of the BdG Hamiltonian in Sec.~\ref{sec:ModeforBdG}, including zero modes and generalized eigenmodes. We show that these modes can be used for constructing the complete solution of the BdG equation. Finally, in Sec.~\ref{sec:Quantisation}, we quantise the BdG field solution and calculate the expectation value of photon number density and anomalous photon number density relevant for section~\ref{sec:Classical}.
\subsection{Mathematical properties of the Bogoliubov-de Gennes Hamiltonian}
\label{sec:MathBdG}
With $\Delta\hat{E}(t,x)=\hat{\psi}(t,x)\e^{it/2}$ and $\tilde{E}_\s :=\sech x$ the BdG equation becomes:
\begin{align}
\left(i\partial_t+\frac{\partial_x^2}{2}+2\tilde{E}_\s^2-\frac{1}{2}\right)\hat{\psi}+ \tilde{E}_\s^2\hat{\psi}^\dagger=0.\label{eq:BdG}
\end{align}
We are interested in the mode solutions of ~\eqref{eq:BdG}. Thus, for simplicity we revert $\psi$ back to a classical field.  
Let us define the Nambu spinor $|\Phi\rangle:=(\psi,\psi^*)^{\intercal}$ in order to rewrite the BdG equation as a Schr\"odinger equation:
\begin{equation}\label{eq:bogo}
    (i\partial_t-\pmb{H}_\BdG)|\Phi\rangle = 0,
\end{equation}
where
\begin{equation}\label{eq:BdGHamiltonian}
\pmb{H}_\BdG:=\left(-\frac{\partial_x^2}{2}-2\tilde{E}_s^2+\frac{1}{2}\right)\sigma_3-i\sigma_2\tilde{E}_s^2
\end{equation}
is the BdG Hamiltonian.
The matrices $\sigma_i$, $i\in\{1,2,3\}$ denote the Pauli matrices.
With a linear Hermitian automorphism operator $\eta$ for a space of operators $\pmb{O}$ on the Hilbert space, one can define the $\eta$-pseudo-Hermitian adjoint
\begin{equation}
    \label{eq:pseudohermitianAdjoint}
    \pmb{O}^\sharp:=\eta^{-1}\pmb{O}^{\dagger}\eta.
\end{equation}
Hence, if $\pmb{O}$ is identical to the pseudo-Hermitian adjoint of the operator, $\pmb{O}^\sharp =\pmb{O}$, it is called $\eta$-pseudo-Hermitian \cite{Mostafazadeh2002,Mostafazadeh2002II,Mostafazadeh2002III,Mostafazadeh2004}.
From Eq.~\eqref{eq:BdGHamiltonian}, one can easily show that the BdG Hamiltonian $\pmb{H}_\BdG$ is $\sigma_3$-pseudo-Hermitian, 
\begin{equation}\label{eq:pseudohermitian}
    \pmb{H}_\BdG^\sharp=\sigma_3^{-1}\pmb{H}_\BdG^{\dagger}\sigma_3 = \pmb{H}_\BdG.
\end{equation}
Since $\sigma_3^{-1}=\sigma_3^\dagger=\sigma_3$, the $\sigma_3$-pseudo-adjoint is not only a similarity transform, but also a unitary transform.
The $\sigma_3$-pseudo-Hermiticity of the Hamiltonian leads to a conserved inner product \cite{Mostafazadeh2010,Mostafazadeh2013,Bender2003,Bender2008}
\begin{equation}
    \langle\Phi|\Phi\rangle_{\sigma_3} := \langle \Phi|\sigma_3|\Phi\rangle.
\end{equation}

Another property of the $\pmb{H}_\BdG$ is the so-called particle-hole symmetry
\begin{equation}\label{eq:ptlholesym}
    \sigma_1\pmb{H}_\BdG\sigma_1=-\pmb{H}_\BdG^*.
\end{equation}
It can be used to find a negative frequency (or negative norm) partner mode of a positive frequency mode of Eq.~\eqref{eq:bogo} \footnote{In fact, $\pmb{H}_\BdG^*=\pmb{H}_\BdG$ here, but we use the complex conjugate for to conform to literature~\cite{LucaThesis}.}. 
To show this, let $|\Phi^{(\omega)}\rangle$ be the normalised positive norm right eigenmode of the BdG Hamiltonian satisfying
\begin{equation}\label{eq:eigenvalueeq}
    \pmb{H}_\BdG|\Phi^{(\omega)}\rangle = \omega|\Phi^{(\omega)}\rangle
\end{equation}
and
\begin{equation}\label{eq:FieldOperatorNormalization}
    \langle \Phi^{(\omega)}|\Phi^{(\omega)}\rangle_{\sigma_3} = 1.
\end{equation}
It is known that ${\rm Spec}(\pmb{H_{\BdG}})\subset\mathbb{R}$ and therefore, $\omega^*=\omega$.
Using Eqs.~\eqref{eq:ptlholesym} and \eqref{eq:eigenvalueeq}, one can show that
\begin{equation}\label{eq:ptlholesymnegative}
    \pmb{H}_\BdG\sigma_1|\Phi^{(\omega)}\rangle^*  = -\omega\sigma_1|\Phi^{(\omega)}\rangle^*.
\end{equation}

Therefore, $|\Bar{\Phi}^{(\omega)}\rangle :=\sigma_1|\Phi^{(\omega)}\rangle^* $ is a right eigenvector with eigenvalue $-\omega$.
Moreover, the spinor has negative norm as
\begin{equation}
    \langle \Phi^{(\omega)}|\sigma_1\sigma_3\sigma_1|\Phi^{(\omega)}\rangle = -\langle \Phi^{(\omega)}|\sigma_3|\Phi^{(\omega)}\rangle = -1.
\end{equation}
Positive and negative norm states appear in pairs, giving the property \eqref{eq:ptlholesym} its name, particle-hole symmetry. 
Like for Hermitian operators, the eigenstates for differing eigenvalues are orthogonal.
Using a complete set of eigenstates, the field can be expanded in the form
\begin{align}
    |\Phi\rangle &= \SumInt |\Phi^{(\omega)}\rangle\langle \Phi^{(\omega)}|\Phi\rangle_{\sigma_3},
\end{align}
where $\SumInt$ is the symbolic expression for summing and integrating over all possible modes. 
Because  we now have a formal expansion for the spinor field $|\Phi\rangle$, we need to find a complete set of modes $|\Phi^{(\omega)}\rangle$.

\subsection{Mode Structure for the Bogoliubov-de Gennes Hamiltonian}
\label{sec:ModeforBdG}
We obtain the eigenmodes of the BdG Hamiltonian by the inverse scattering transform method \cite{Yang2000}. 
With $|\Phi^{(\omega)}(t,x)\rangle:=|\Phi_k(x)\rangle\e^{-i\omega t}$, Eq.~\eqref{eq:bogo} becomes the mode equation
\begin{equation}
    (\omega-\pmb{H}_\BdG)|\Phi_k\rangle=0,
\end{equation}
with explicit positive frequency modes
\begin{equation}\label{eq:ContinuousSpinor}
    |\Phi_k\rangle=\frac{1}{\sqrt{2\pi}(k^2+1)}\begin{pmatrix}
        (\tanh{x}+ik)^2\\-\sech^2\!\!x
    \end{pmatrix}\e^{-ikx},
\end{equation}
where the eigenmodes satisfy
\begin{equation}
    \pmb{H}_\BdG|\Phi_k\rangle =\omega_k|\Phi_k\rangle
\end{equation}
and the eigenfrequencies are
\begin{equation}
    \omega_k=\frac{1+k^2}{2}.
    \label{eq:eigenfrequencies}
\end{equation}
Using the particle-hole symmetry, we find the eigenmodes with a negative eigenvalue by using Eq.~\eqref{eq:ptlholesymnegative},
\begin{equation}
    |\Bar{\Phi}_k\rangle := \sigma_1|\Phi_k\rangle^*.
\end{equation}
The negative frequency partner modes have the same frequencies as in relation \eqref{eq:eigenfrequencies}, but of opposite sign.
Overall, the states satisfy the orthonormality relations:
\begin{equation}\label{eq:ContinuousModeOrthonornality}
    \langle \Phi_k|\Phi_{k'}\rangle_{\sigma_3} = -\langle \Bar{\Phi}_k|\Bar{\Phi}_{k'}\rangle_{\sigma_3} = \delta (k-k'),
 \qquad \langle \Phi_k|\Bar{\Phi}_{k'}\rangle_{\sigma_3} = 0.
\end{equation}
The modes are best labelled by $k$, as each real $k$ indicates one mode pair.

The eigenmodes we have found at this point have either positive or negative, i.e. nonzero continuous real eigenvalues. These, however, do not form a complete basis of the solution space  \cite{Yang2000}.  One reason for this is that the inverse scattering method does only describe modes that are asymptotic plane waves.
In fact, we need four more modes to complete the basis, because the Hamiltonian~\eqref{eq:BdGHamiltonian} is a complex second order differential operator. Two eigenmodes can be found using the eigenvalue zero, which do not correspond to plane waves. We call them zero modes: 
\begin{align}
    |\Phi_{(1)}^\zero\rangle&=\frac{1}{\sqrt{2}}\sech x \begin{pmatrix}
        1\\-1
    \end{pmatrix},\\
    |\Phi_{(2)}^\zero\rangle&=\frac{1}{\sqrt{2}}\tanh{x} \,\, \sech x \begin{pmatrix}
        1\\1
    \end{pmatrix}.
\end{align}
Now we are looking for the two further modes to complete the space. Note that the zero modes have zero norm. For the continuous modes, the particle-hole symmetry allowed us to find negative norm partner modes. The zero modes are their own partner modes in the particle-hole symmetry. Therefore, we need to find the two further modes by other means.
It is known that each zero mode is accompanied by a `generalised eigenmode' \cite{Ripka1986}. These generalised eigenmodes are not eigenstates of the BdG Hamiltonian but can be used to find further linearly independent solutions of the BdG equation.
\begin{align}
    |\Phi_{(1)}^\GE\rangle&=\frac{1}{\sqrt{2}}(x\tanh{x}-1)\sech x \begin{pmatrix}
        1\\1
    \end{pmatrix},\\
    |\Phi_{(2)}^\GE\rangle&=\frac{1}{\sqrt{2}}x\sech x \begin{pmatrix}
        1\\-1
    \end{pmatrix}.
\end{align}
These modes satisfy
\begin{align}
    \pmb{H}_\BdG|\Phi_{(i)}^\GE\rangle = |\Phi_{(i)}^\zero\rangle, \quad i=1,2.
\end{align}
The generalised eigenmodes have zero norm, too. The existence of these modes  is a consequence of the nondiagonalizability of the BdG Hamiltonian \cite{Bender2008,Sokolov2006}. 
The zero modes and generalised eigenmodes both have zero norm.
Moreover, they are orthogonal to the continuous modes. 
The nonzero inner products are
\begin{align}
    \langle \Phi_{(i)}^\zero|\Phi_{(j)}^\GE\rangle_{\sigma_3} &= \langle\Phi_{(i)}^\GE|\Phi_{(j)}^\zero\rangle_{\sigma_3} = (-1)^{i}\delta_{ij}.
\end{align}
From this, we construct further two solutions of the Bogoliubov equation Eq.~\eqref{eq:bogo}:
\begin{equation}
    |\Phi_{(i)}^\col\rangle = -it|\Phi_{(i)}^\zero\rangle + |\Phi_{(i)}^\GE\rangle.
\end{equation}
Since these are independent solutions of the BdG equation, they can be used in the field expansion. Combined with the continuous modes and the zero modes, the field can be expressed as
\begin{align}\label{eq:fieldexpansion}
    |\Phi\rangle &= \int\ud k  \big(a_k \e^{-i\omega_k t}|\Phi_k\rangle+a_k^*\e^{i\omega_k t} |\Bar{\Phi}_k\rangle \big)+\sum_{i=1,2}\big(b_i|\Phi_{(i)}^\zero\rangle +c_i|\Phi_{(i)}^\col\rangle\big)
\end{align}
where $a_k,a_k^*,b_i,c_i\in\mathbb{C}$. The coefficient of each negative eigenvalue mode, $a_k^*$, is chosen to satisfy the particle-hole symmetry. 
Importantly, the norm of this field is conserved:
\begin{equation}
    \partial_t\langle\Phi|\Phi\rangle_{\sigma_3}=0.
\end{equation}
In addition, any initial field at $t=0$  can be expressed using the expansion by virtue of the completeness relation 
\begin{align}\label{eq:Completeness}
    \mathds{1}_2&\delta(x-x')\nonumber\\
    &=\int \ud k\big(|\Phi_k(x)\rangle\langle\Phi_k(x')|\sigma_3+|\Bar{\Phi}_k(x)\rangle\langle\Bar{\Phi}_k(x')|\sigma_3\big)+\sum_{i}|\Phi_{(i)}^\zero(x)\rangle\langle\Phi_{(i)}^\col(x')|\sigma_3+|\Phi_{(i)}^\col(x)\rangle\langle \Phi_{(i)}^\zero(x')|\sigma_3,
\end{align}
where $\mathds{1}_2$ is the $2\times2$ identity matrix.

Finally, we mention a relation between the discrete modes and the linear perturbations of the soliton parameters, which can be found in  \cite{Haus1990}.
The NLS soliton, written in soliton parameters, writes:
\begin{equation}
    E_\s = n_0\e^{i(n_0^2-p_0^2)t/2}\sech[n_0(x-x_0)-n_0p_0t]\e^{ip_0(x-x_0)}\e^{i\theta_0}.
\end{equation}
where $n_0$ is the soliton photon number, $x_0$ the soliton position, $\theta_0$ the soliton phase, and $p_0$  the soliton momentum.
Each of the four discrete modes is related to a variation of a soliton parameter in the following way:
\begin{align}
    v_n &= \partial_{n_0}E_\s = -\pmb{P}_1|\Phi^\col_{(1)}\rangle,\\
    v_\theta &= \partial_{\theta_0}E_\s=i\pmb{P}_1|\Phi_{(1)}^\zero\rangle,\\
    v_p &= \partial_{p_0}E_\s = i\pmb{P}_1|\Phi^\col_{(2)}\rangle,\\
    v_x &= \partial_{x_0}E_\s = \pmb{P}_1|\Phi_{(2)}^\zero\rangle,
\end{align}
where $\pmb{P}_1=(\mathds{1}_2+\sigma_3)/2$ is the projection operator to the first row. 
This identification shows that the modes that are related with the soliton parameter perturbations or, equivalently, the collective modes are the discrete modes, which are necessary to complete the basis expansion of the field.  We observe that the perturbations corresponding to the zero modes, $\theta_0$ and $x_0$, are conserving the energy, whereas perturbations of the collective modes, $n_0$ and $p_0$ , are conjugate to them and incrementally change the soliton energy.
In Sec.~\ref{sec:Quantisation}, we now present the quantization scheme and define the vacuum configuration.
Unlike the continuous modes, the discrete modes are not of the plane wave form, even far from the soliton.
Therefore, discrete modes cannot be interpreted as a quasi-particles. 

\subsection{Quantisation of the field operator}
\label{sec:Quantisation}
We proceed now to quantise the field expansion,  Eq. \eqref{eq:fieldexpansion}. The expansion of the quantum field is obtained by replacing the classical mode amplitudes with mode operators that satisfy the commutation relations 
\begin{equation}
    [\hat{a}_k,\hat{a}_{k'}^\dagger] = \delta(k-k')
\end{equation}
and with 
\begin{equation}
    n:=-c_1 \quad \theta:=-i\,b_1 \quad p:=-i\,c_2 \quad x:=b_2,
\end{equation}
\begin{equation}
    [\hat{\theta},\hat{n}]= [\hat{x},\hat{p}]=i.
\end{equation}
All other pairs of operators commute. 
Now, the quantised Bogoliubov-field spinor $|\hat{\Phi}\rangle:=(\hat{\psi},\hat{\psi}^\dagger)^\intercal$ can be written in the form,
\begin{align}
    |\hat{\Phi}\rangle &= \int \ud k \big(\hat{a}_k\e^{-i\omega_k t}|\Phi_k\rangle+\hat{a}_k^\dagger\e^{i\omega t}|\Bar{\Phi}_k\rangle\big)+i\hat{\theta}|\Phi_{(1)}^\zero\rangle+\hat{x}|{\Phi}_{(2)}^\zero\rangle-\hat{n}|\Phi_{(1)}^\col\rangle+i\hat{p}|{\Phi}_{(2)}^\col\rangle\big).
    \label{eq:modeexpansion}
\end{align}
Before moving on, it is useful to rewrite the expansion in terms of  creation and  annihilation operators  also for the discrete modes. 
Recall that the expectation value of the quantum field is zero, implicit in Eq.~\eqref{eq:electricfieldexpectation}, 
\begin{equation}\label{eq:electricfieldvev}
    \langle 0 | \hat{\psi} |0\rangle = 0,
\end{equation}
where the vacuum state $|0\rangle$ is defined at $t=0$ and we work in the Heisenberg picture. Hence, expectation values are more easily calculated if also the discrete modes in the expansion of $\hat{\psi}$ are represented by annihilation and creation operators.
We define these operators using the soliton parameter operators:
\begin{equation}
    \hat{\alpha}_{(1)} = \frac{1}{\sqrt{2}}(\hat{\theta}+i\hat{n}),\qquad \hat{\alpha}_{(2)} = \frac{1}{\sqrt{2}}(\hat{x} + i\hat{p}).
\end{equation}
which satisfy the discrete bosonic commutation relation $[\hat{\alpha}_{(i)},\hat{\alpha}^\dagger_{(j)}]=\delta_{ij}$ and commute with all the creation/annihilation operators of the continuous modes.
In contrast to $\hat{a}_k$ and $\hat{a}^\dagger_k$ in Eq.~\eqref{eq:modeexpansion}, these operators do not represent the creation of quasi-particles, i.e. asymptotic plane wave solutions. Although the excitations do not come as particle-antiparticle pairs, we can consistently define the vacuum state with:
\begin{equation}
    \hat{\alpha}_{(i)}|0\rangle = 0,
\end{equation}
which is required by Eq.~\eqref{eq:electricfieldvev}.
In this representation, the field expansion writes:
\begin{align}
    |\hat{\Phi}\rangle &= \int \ud k \big(\hat{a}_k\e^{-i\omega_k t}|\Phi_k\rangle+\hat{a}_k^\dagger\e^{i\omega_k t}|\Bar{\Phi}_k\rangle\big)+\sum_{i\in\{1,2\}}\big(\hat{\alpha}_{(i)}|\Phi_{(i)}\rangle+\hat{\alpha}_{(i)}^\dagger|\Bar{\Phi}_{(i)}\rangle\big),\label{eq:BogoliobovSpinorExpansion}
\end{align}
where
\begin{align}
    |\Phi_{(1)}\rangle &= \frac{i}{\sqrt{2}}\big(|\Phi_{(1)}^\zero\rangle+|\Phi_{(1)}^\col\rangle\big),\label{eq:DiscreteSpinor1}\\
    |\Bar{\Phi}_{(1)}\rangle&= \sigma_1|{\Phi}_{(1)}\rangle^* = -\frac{i}{\sqrt{2}}\big(|\Phi_{(1)}^\zero\rangle-|\Phi^\col_{(1)}\rangle\big),\\
    |\Phi_{(2)}\rangle&=\frac{1}{\sqrt{2}}\big(|\Phi_{(2)}^\zero\rangle+|\Phi^\col_{(2)}\rangle\big),\label{eq:DiscreteSpinor2}\\
    |\Bar{\Phi}_{(2)}\rangle&=\sigma_1|{\Phi}_{(2)}\rangle^*=\frac{1}{\sqrt{2}}\big(|\Phi_{(2)}^\zero\rangle-|\Phi^\col_{(2)}\rangle\big).
\end{align}
The discrete modes introduced here have a nonzero norm. 
For zero modes, we cannot use the property of particle-hole symmetry, Eq. \eqref{eq:ptlholesymnegative}, to find the partner mode \cite{Baak2022}. But combinations of these states satisfy Eq. \eqref{eq:ptlholesymnegative} and restore the partnered relation analogous to the continuous mode.
At $t=0$, these modes form an orthonormal set similar to the continuous modes (Eq.~\eqref{eq:ContinuousModeOrthonornality}),
\begin{equation}\label{eq:DiscreteModeOrthonormality}
    \langle \Phi_{(i)}|\Phi_{(j)}\rangle_{\sigma_3} = -\langle \Bar{\Phi}_{(i)}|\Bar{\Phi}_{(j)}\rangle_{\sigma_3} = \delta_{i,j}, \qquad \langle \Phi_{(i)}|\Bar{\Phi}_{(j)}\rangle_{\sigma_3}=0. 
\end{equation}

 Recall that $|\hat{\Phi}\rangle=(\hat{\psi},\hat{\psi}^\dagger)^\intercal$  and $(u_k,v_k)^\intercal=|\Phi_k(x)\rangle\e^{-i\omega t}$.
The quantised Bogoliubov expansion \eqref{eq:BogoliobovSpinorExpansion} in terms of mode functions is
\begin{align}
    \hat{\psi}(t,x) &= \int \ud k \big(\hat{a}_ku_k(t,x)+\hat{a}_k^\dagger v_k^*(t,x)\big) + \sum_{i\in\{1,2\}} \big( \hat{\alpha}_{(i)} u_{(i)}(t,x) +\hat{\alpha}_{(i)}^\dagger v_{(i)}^*(t,x)\big),
\end{align}
with a similar expression for $\hat{\psi}^\dagger$. We now have a suitable set of modes to calculate expectation values. The amplitudes of the modes, the mode functions $(u_k,v_k)^\intercal$, can now be calculated.[$|\Phi^{(\omega)}(t,x)\rangle:=(u_k,v_k)^\intercal:=|\Phi_k(x)\rangle\e^{-i\omega t}$] The discrete modes are defined in analogy to the continuous modes, $|\Phi_{(i)}\rangle := (u_{(i)},v_{(i)})^\intercal$.
We derive the explicit form of the mode function and calculate the density $\langle\hat{\psi}^\dagger(t,x)\hat{\psi}(t,x)\rangle$ and anomalous density $\langle\hat{\psi}^2(t,x)\rangle$, as these will appear in the backreaction field $E_c$.
In what follows, we separately treat the contributions arising  from the discrete and continuous modes, respectively.
Total densities are simply a sum of these contributions, because discrete and continuous modes commute and there is no mixture of discrete and continuous mode contribution in the vacuum expectation value of the second-order correlations.
We start with the continuous mode contribution.
The mode functions can be obtained from Eq.~\eqref{eq:ContinuousSpinor}:
\begin{align}
    u_k &= \frac{1}{\sqrt{2\pi}(k^2+1)}(\tanh{x}+ik)^2\e^{-ikx}\e^{-i\omega_k t},\\
    v_k^* &= -\frac{1}{\sqrt{2\pi}(k^2+1)}\sech^2\!\!x\,\e^{ikx}\e^{i\omega_k t}.
\end{align}
Hence, using the subscript $\con$ for the continuous mode contribution, we calculate:
\begin{align}\label{eq:BogoliubovDensity1}
    \langle\hat{\psi}^\dagger\hat{\psi}\rangle_\con & = \int \ud k \,|v_k|^2=\frac{1}{2\pi}\sech^4\!\!x\int\ud k\frac{1}{(k^2+1)^2}\nonumber\\
    &=\frac{1}{4}\sech^4\!\!x,\\
    \langle\hat{\psi}^2\rangle_\con & =\int \ud k u_kv_k^*= -\frac{1}{2\pi}\sech^2\!\!x\int\ud k \frac{(\tanh x+ik)^2}{(k^2+1)^2}\nonumber\\
    &=-\frac{1}{4}\sech^4\!\!x.\label{eq:BogoliubovDensity2}
\end{align}    
Note that these are real functions and thus  $\langle\hat{\psi}^{\dagger2}\rangle_\con = \langle\hat{\psi}^{2}\rangle_\con$ and $ \langle\hat{\psi}\hat{\psi}^\dagger\rangle_\con= \langle\hat{\psi}^\dagger\hat{\psi}\rangle_\con$.
In a similar manner, the discrete modes can be obtained from \eqref{eq:DiscreteSpinor1} and \eqref{eq:DiscreteSpinor2} and the explicit solutions given in Sec.~\ref{sec:ModeforBdG}:
\begin{align}
    u_{(1)} &=\frac{1}{2}(ix\tanh{x}+t)\sech{x},\label{eq:DiscretePositiveMode1}\\
    v_{(1)} &=\frac{1}{2}\big(i(x\tanh{x}-2)-t\big)\sech{x},\label{eq:DiscreteNegativeMode1}\\
    u_{(2)} &=\frac{1}{2}\big((1-it)\tanh{x}+x\big)\sech{x},\label{eq:DiscretePositiveMode2}\\
    v_{(2)} &=\frac{1}{2}(\big(1-it)\tanh{x}-x\big)\sech{x}\label{eq:DiscreteNegativeMode2}.
\end{align}
Finally, using the subscript $\dis$ for the discrete mode contribution, we obtain:
\begin{align}\label{eq:BogoliubovDensity3}
    \langle\hat{\psi}^\dagger\hat{\psi}\rangle_\dis &= \sum_i|v_{(i)}|^2\nonumber\\
    &= \frac{1}{4}\big[\big((1+x^2)\tanh^2\!\!{x}-6x\tanh{x}+x^2+4\big)+t^2(2-\sech^2\!\!{x})\big]\sech^2\!\!x,\\
    \langle\hat{\psi}^2\rangle_\dis &= \sum_iu_{(i)}v_{(i)}^*\nonumber\\
    &= \frac{1}{4}\big[\big((1+x^2)\tanh^2\!\!{x}-2x\tanh{x}-x^2\big)+2 i t-t^2 \sech^2\!\!x\big]\sech^2\!\!x. \label{eq:BogoliubovDensity4}
\end{align}
Here, $\langle\hat{\psi}\hat{\psi}^{\dagger}\rangle_\dis =  \langle\hat{\psi}^\dagger\hat{\psi}\rangle_\dis$ and $\langle\hat{\psi}^{\dagger2}\rangle_\dis = \langle\hat{\psi}^{2}\rangle_\dis^*$.
Overall the second-order correlations of the continuous modes lack time dependence, whereas the discrete mode correlations increase quadratically in time. In the next section we show how this quadratic behaviour is also found in the correction field as a result of number conservation. 

\section{The classical field correction from the backreaction}
\label{sec:Classical}
In the previous section, we solved the Bogoliubov-de Gennes equation and calculated the relevant expectation values.
In this section, we solve the equation of motion for the correction field $E_\c$, which describes a backreaction.
Let us write $E_c(t,x)=e^{it/2}\tilde{E}_c(t,x)$ and define the spinor $|E_c\rangle = (\tilde{E}_c,\tilde{E}_c^*)^\intercal$, which, because of Eq.~\eqref{eq:GPcorr}, satisfies 
\begin{equation}\label{eq:correction}
    (i\partial_t-\pmb{H}_\BdG)|E_c\rangle =|{\rm source}\rangle ,
\end{equation}
where
\begin{equation}\label{eq:source}
    |{\rm source}\rangle=\begin{pmatrix}
        -\big(2\langle\hat{\psi}^\dagger\hat{\psi}\rangle+\langle\hat{\psi}^2\rangle\big) \tilde E_\s\\
        \big(2\langle\hat{\psi}^\dagger\hat{\psi}\rangle+\langle\hat{\psi}^{\dagger2}\rangle\big) \tilde E_\s
    \end{pmatrix}.
\end{equation}
Note that, apart from the inhomogeneous source term, the equation is the same as the BdG equation (compare also the left-hand sides of Eqns.~\eqref{eq:GPcorr} and ~\eqref{eq:BdG}).
Consequently, the classical field correction can be written in the form
\begin{align}\label{eq:correctionfieldExpansion}
    E_\c&=\e^{it/2}\Bigg[\int\ud k \big(b_ku_k+b_k^*v_k^*\big) + \sum_i \big(c_{(i)}u_{(i)} +c_{(i)}^{*}v_{(i)}^*\big)\Bigg]+E_\c^{(p)}.
\end{align}
where superscript ${(p)}$ is used to denote a particular solution. 
$b_k$ and $c_{(i)}$ are complex numbers determined by the initial condition. 
The solution of Eq. \eqref{eq:correction} starts with a particular solution.
The source term \eqref{eq:source} only contains the density and anomalous density whose continuous and discrete mode contributions we already calculated in section \ref{sec:Quantisation}. 
Since the operator $(i\partial_t-\pmb{H}_\BdG)$ is a linear operator, we can split the particular solution into the contributions from  the continuous and the discrete mode densities, respectively:
\begin{equation}
    E_\c^{(p)} = E_{\c,\con}^{(p)}+E_{\c,\dis}^{(p)}.
\end{equation}

We start with the continuous contribution. A detailed calculation of what is summarized below is given in App.~\ref{App:CorrectionCalculation}.
Using Eqns.~\eqref{eq:BogoliubovDensity1} and \eqref{eq:BogoliubovDensity2}, the part of the source \eqref{eq:source}  originating from the continuous mode densities is
\begin{equation}
    |{\rm source}\rangle_\con=-\frac{1}{4}\sech^5x\begin{pmatrix}
        1\\-1
    \end{pmatrix}.
\end{equation}
A particular solution for the continuous contribution is ~\eqref{eq:conparticular}:
\begin{equation}
    E_{\c,\con}^{(p)}(t,x)=(f_{0,\con}(x)+itg_{1,\con}(x))\e^{it/2},
\end{equation}
where
\begin{align}
    g_{1,\con}(x)&=-\frac{1}{9}\sech{x},\label{eq:g1Cont}\\
    f_{0,\con}(x)&=\frac{1}{36} (3\sech^2\!\!x-10+4x \tanh{x} )\sech{x}.\label{eq:f0Cont}
\end{align}

For the discrete mode source term we also insert Eq.~\eqref{eq:BogoliubovDensity3} and Eq.~\eqref{eq:BogoliubovDensity4} into Eq.~\eqref{eq:source},
\begin{equation}
    |{\rm source}\rangle_\dis=S_R(t,x)\begin{pmatrix}
        1\\-1
    \end{pmatrix}+iS_I(t,x)\begin{pmatrix}
        1\\1
    \end{pmatrix}.
\end{equation}
where 
\begin{align}\label{eq:DiscreteRealSource}
    S_R(x)&:=-\frac{1}{4}\big(8+x^2-14x\tanh{x}+3(x^2+1)\tanh^2\!\!x\big)\sech^3\!\!x-\frac{1}{4}t^2(1+3\tanh^2\!\!x)\sech^3\!\!{x},\\
    S_I(x)&:=-\frac{1}{2} t\sech^3x.\label{eq:DiscreteImaginarySource}
\end{align}
The particular solution that vanishes in the asymptotic region $|x|\to\infty$ is ~\eqref{eq:disparticular} 
\begin{equation}
    E_{\c,\dis}^{(p)} = \big(f_{2,\dis}(x)t^2+f_{0,\dis}(x)\big)\e^{it/2}.
\end{equation}
where 
\begin{align}
    f_{2,\dis}(x)&=-\frac{1}{4} \sech^3x,\label{eq:f2Dis}\\
    f_{0,\dis}(x)&=\frac{4}{15}\Big(x \sinh{x}+\cosh{x}\big(\ln{(\sech{x})}-\ln2\big)\Big)-\frac{1}{4} (1+x^2)\sech^3\!\!{x}\nonumber\\
    &\quad -\frac{1}{60}\big(41+48\ln2+48\ln (\sech{x})\big)\sech{x}\nonumber\\
   &\quad -\frac{2}{15} \Big(6 \, \text{Li}_2\big(-e^{-2 x}\big)+x \big(6 x+6 \ln (\sech{x})+5-6 \ln{2}\big)\Big)\tanh{x}\sech{x}\label{eq:f0Dis}
\end{align}
where $\mathrm{Li}_s(z)$ is a polylogarithm (Jonqui\`ere's function).
Because the inhomogeneity of Eq.~\eqref{eq:correction} is well behaved and vanishes asymptotically, physical solutions of this equation are described in the form of  Eq.~\eqref{eq:correctionfieldExpansion} .

We further point out a $t^2$ contribution to $E_\dis^{(p)}$ , which is not present in the  Bogoliubov field solutions derived at the end of the previous section.
Therefore, this quadratic dependence is a characteristic of the backreaction effect via the discrete mode contribution $E_\dis^{(p)}$. It is independent of the initial condition and, hence,  in the following we will be able to identify quadratic terms caused by the backreaction. 
In other words, the quadratic decay of the soliton photon number investigated in \cite{Ward2023} is in fact the backreaction effect from the discrete mode.

We now have derived an analytic form of the general solution of the backreaction field $E_c$. 
\section{Exploration of Results}
\label{sec:Physical}
Here, we will explore the resultant correction field solutions further.
Since we know the mode expansion of  the Bogoliubov field and  the correction field, it is possible to calculate  physical quantities.
As an example, we calculate the photon number density and photon number of the Bogoliubov and correction fields, respectively.
We recover the units in order to assess the size of these fields.
The units can be recovered by putting dimensional constants in each dimensionless parameters.
For the soliton, 
\begin{equation}
    E_\s = \sqrt{P_0}\sech(\tau/T_0)\e^{i\zeta/2L_D}.
\end{equation}
where $P_0$ is the soliton peak power, $T_0$ is the temporal width and $L_D$ is the dispersion length
\begin{equation}
    L_D=\frac{T_0^2}{|k_2|}.
\end{equation}
where $k_2$ is the group velocity dispersion parameter at the soliton carrier frequency \cite{Agrawal2019}.
In our number-conserving approach, the expansion parameter was identified as $\epsilon=1/\sqrt{n}_0$.
Hence, physical dimensions are recovered by identifying $t\equiv \zeta/L_D$ and $ x\equiv\tau/T_0$.
Also, this  relates to the  total soliton photon number by
\begin{equation}
    2P_0T_0 = n_0\hbar\bar{\omega}_\s
\end{equation}
where $\bar{\omega}_\s$ is the average frequency of the soliton.

\subsection{Photon number density}
\label{sec:numberdensity}
Since we have expanded the field in orders related to the photon number, the photon number density is a useful physical quantity to explore.
In this section, we explore the expectation value of  the photon number density of each of the fields in \eqref{eq:orderexpansion},  focusing on long propagation distances $(\zeta\gg L_d)$.
We also consider the total photon number of each field separately to monitor the number conservation, treating continuous and discrete mode contributions separately.
We start with the photon number density expectation value, which can be written with Eq.~\eqref{eq:orderexpansion} in the form:
\begin{equation}
    \rho:=\langle \hat{\rho}\rangle = |E_\s|^2 + \epsilon^2\langle \hat{\psi}^\dagger\hat{\psi}\rangle + 2\epsilon^2\Re[E_\s^*E_\c] +\mathcal{O}(\epsilon^3) := \rho_\s+\rho_{\Delta}+\rho_\c+\mathcal{O}(\epsilon^3),
\end{equation} 
where $\rho_\s$ is the photon number density from the soliton background, $\rho_\Delta$ is the Bogoliubov field contribution to the photon number density, and $\rho_\c$ is the photon number density from the backreaction effect. Note that this is a density in time, because the QNLS describes the evolution in space.
Also note that both $\rho_\Delta$ and $\rho_\c$ are of the same order ($\epsilon^2$). 
For the Bogoliubov field, the  photon number density from continuous and discrete modes, respectively, were obtained earlier as Eq.~\eqref{eq:BogoliubovDensity1} and Eq.~\eqref{eq:BogoliubovDensity3}.
From these, the photon number density  $\rho_{\Delta}^\con$ for continuous  and $\rho_{\Delta,\dis}$ for discrete modes  can be obtained in physical dimensions of inverse time:
\begin{align}
    \rho_{\Delta,\con} &= \frac{1}{2T_0}\langle\hat{\psi}^\dagger\hat{\psi}\rangle_\con = \frac{1}{8T_0}\sech^4(\tau/T_0),\\
    \rho_{\Delta,\dis} &= \frac{1}{2T_0}\langle\hat{\psi}^\dagger\hat{\psi}\rangle_\dis = \frac{1}{8T_0}\Big[5+2(\tau/T_0)^2-6(\tau/T_0)\tanh{(\tau/T_0)}-\big(1+(\tau/T_0)^2\big)\sech^2\!\!{(\tau/T_0)}\nonumber\\
    & \qquad\qquad\qquad\qquad\qquad\quad+(\zeta/L_D)^2 \big(2-\sech^2\!\!{(\tau/T_0)}\big)\Big]\sech^2(\tau/T_0).
\end{align}
For long propagation distances, the  quadratic $\zeta$-dependent term of the discrete mode dominates,
\begin{equation}
    \rho_\Delta:=\rho_{\Delta,\con}+\rho_{\Delta,\dis}\xrightarrow[]{\zeta\gg L_D}\frac{1}{8T_0}(\zeta/L_D)^2\big(2-\sech^2(\tau/T_0)\big)\sech^2
    (\tau/T_0).
\end{equation}
As the soliton propagates, the Bogoliubov field photon number density increases .
In comparison we consider the photon number density of the backreaction field, or equivalently, the classical field correction.
We will also treat the continuous- and discrete mode contributions, $\rho_{\c,\con}$ and $\rho_{\c,\dis}$, separately.  Again, using a correction field without the homogeneous part, we obtain:
\begin{align}
    \rho_{\c,\con}&=\frac{1}{2T_0}2\Re[E_\s^*E_{\c,\con}]=\frac{1}{36T_0} \big(3\sech^2(\tau/T_0)-10+4(\tau/T_0) \tanh{(\tau/T_0)} \big)\sech^2{(\tau/T_0)}.\\
    \rho_{\c,\dis}&=\frac{1}{2T_0}2\Re[E_\s^*E_{\c,\dis}]=\frac{4}{15T_0}\Big((\tau/T_0) \sinh{(\tau/T_0)}+\cosh{(\tau/T_0)}\big(\ln{(\sech{(\tau/T_0)})}-\ln2\big)\Big)\sech\!{(\tau/T_0)}\nonumber\\
    &\qquad\qquad\qquad\qquad\quad-\frac{1}{4T_0} \big(1+(\tau/T_0)^2\big)\sech^4\!{(\tau/T_0)} -\frac{1}{60T_0}\big(41+48\ln2+48\ln (\sech{(\tau/T_0)})\big)\sech^2\!{(\tau/T_0)}\nonumber\\
   &\qquad\qquad\qquad\qquad\quad -\frac{2}{15T_0} \Big((\tau/T_0) \big(6 (\tau/T_0)+6 \ln (\sech{(\tau/T_0)})+5-6 \ln{2}\big)\Big)\tanh{(\tau/T_0)}\sech^2{(\tau/T_0)}\nonumber\\
    &\qquad\qquad\qquad\qquad\quad-\frac{4}{5T_0} \text{Li}_2\big(-e^{-2 (\tau/T_0)}\big)\tanh{(\tau/T_0)}\sech^2{(\tau/T_0)}\nonumber\\
    &\qquad\qquad\qquad\qquad\quad-\frac{1}{4T_0}(\zeta/L_D)^2\sech^4(\tau/T_0)
\end{align}
Eventually, the quadratic increasing term of $\zeta$ in discrete mode contribution dominates the density:
\begin{equation}
    \rho_\c:=\rho_{\c,\con}+\rho_{\c,\dis}\xrightarrow[]{\zeta\gg L_D}-\frac{1}{4T_0}(\zeta/L_D)^2\sech^4(\tau/T_0).
    \label{eq:quadraticpart}
\end{equation}
The minus sign means that the photon number density of the backreaction reduces the total photon number  density, i.e. it compensates 
the increase in the Bogoliubov field.
Since both photon number densities we calculated here are in the same order of the expansion parameter $(\epsilon^2)$, it is interesting also to consider the sum of both contributions.
Since both of these belong to the soliton, we call the sum of them, $\rho_\Delta+\rho_\c$, the `photon number density fluctuation' of the soliton.
\begin{figure}[ht]
\includegraphics[scale=0.6]{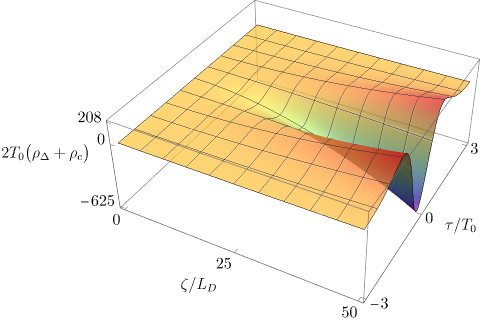}
\caption{Distortion of the soliton shape envelope due to quantum fluctuations. Evolution of the change in photon number density of the soliton. The initial configuration of $E_\c$ is chosen according to Eqs. \eqref{eq:f0Cont}, \eqref{eq:g1Cont}, \eqref{eq:f0Dis}, \eqref{eq:f2Dis} for an arbitrary small initial distortion. By adding a general solution, any physical initial configuration can be realised. As the soliton propagates along $\zeta$, regardless of initial configuration, the unstable mode contribution to the photon number density becomes dominant.
The deviation of the intensity from the classical soliton intensity increases as soliton propagates ($\sim\mathcal{O}(\zeta^2)$)
}
\label{fig:figmain1}
\end{figure}  
Fig.~\ref{fig:figmain1} shows the photon number density distortion $\rho_\Delta+\rho_\c$ of the soliton.
We obtain
\begin{align}
    \rho_\Delta+\rho_\c &=\frac{4}{15T_0} \Big((\tau/T_0) \sinh{(\tau/T_0)} +\cosh{(\tau/T_0)}\big(\ln{(\sech{(\tau/T_0)})}-\ln2\big)\Big)\sech\!{(\tau/T_0)}\nonumber\\
    &\quad-\frac{1}{24T_0} \big(4+9(\tau/T_0)^2\big)\sech^4\!{(\tau/T_0)} -\frac{1}{360T_0}\big(121-90(\tau/T_0)^2+288\ln2+288\ln (\sech{(\tau/T_0)})\big)\sech^2\!{(\tau/T_0)}\nonumber\\
   &\quad -\frac{1}{180T_0} \Big((\tau/T_0) \big(144 (\tau/T_0)+144 \ln (\sech{(\tau/T_0)})+55-144 \ln{2}\big)\Big)\tanh{(\tau/T_0)}\sech^2{(\tau/T_0)}\nonumber\\
    &\quad-\frac{4}{5T_0} \text{Li}_2\big(-e^{-2 (\tau/T_0)}\big)\tanh{(\tau/T_0)}\sech^2{(\tau/T_0)}+\frac{1}{8T_0}(\zeta/L_D)^2\big(2-3\sech^2(\tau/T_0)\big)\sech^2(\tau/T_0)\nonumber\\
    &\xrightarrow[]{\zeta\gg L_D}\frac{1}{8T_0}(\zeta/L_D)^2\big(2-3\sech^2(\tau/T_0)\big)\sech^2(\tau/T_0). \label{eq:densitydistortion}
\end{align}
As $\zeta$ increases, the $\zeta^2$ contribution to the photon number density fluctuation dominates. Therefore, for any reasonable initial condition, the photon number density fluctuation in the long propagation distance limit will be represented by Eq.~\eqref{eq:densitydistortion}.

\begin{figure}[hb]
     \centering
     \begin{subfigure}[hb]{0.3\textwidth}
         \centering
         \includegraphics[width=\textwidth]{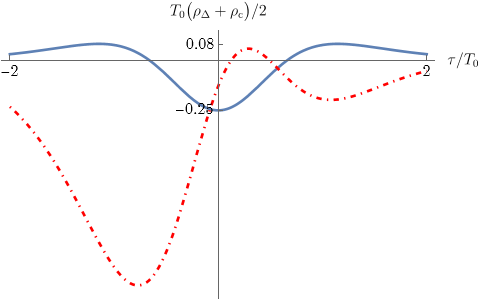}
         \caption{$\zeta/L_D=1$}
         \label{fig:t1graph}
     \end{subfigure}
     \hfill
     \begin{subfigure}[hb]{0.3\textwidth}
         \centering
         \includegraphics[width=\textwidth]{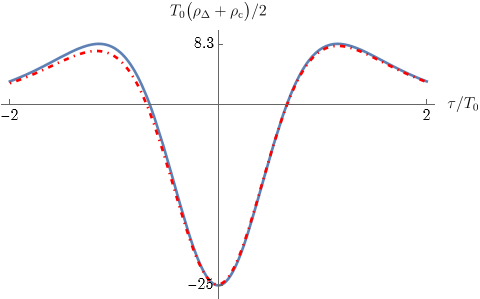}
         \caption{$\zeta/L_D=5$}
         \label{fig:t5graph}
     \end{subfigure}
     \hfill
     \begin{subfigure}[hb]{0.3\textwidth}
         \centering
         \includegraphics[width=\textwidth]{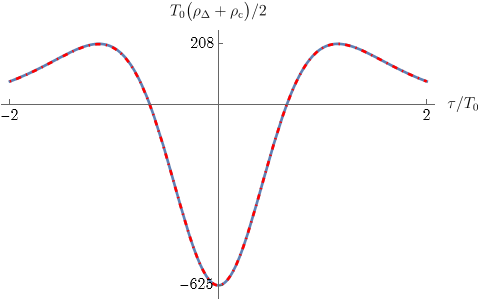}
         \caption{$\zeta/L_D=50$}
         \label{fig:t50graph}
     \end{subfigure}
     \caption{Evolution of the unstable mode in the number density: 
        (a) The total photon number density distortion $\rho_\Delta+\rho_\c$  (red dotted line) might at short distances be arbitrary and large in comparison to its quadratically increasing ($\sim\zeta^2$) contribution (blue line), Eq. \eqref{eq:quadraticpart}, originating from the unstable mode. (b)  After five dispersion lengths, the distortion is almost identical to its unstable mode contribution. (c) The distortion eventually becomes independent of the initial configuration, i.e.
       the deviation always evolves into the characteristic shape of a one-dimensional `crater'.}
        \label{fig:figmain2}
\end{figure}

\begin{figure}
     \centering
     \begin{subfigure}[hb]{0.3\textwidth}
         \centering
         \includegraphics[width=\textwidth]{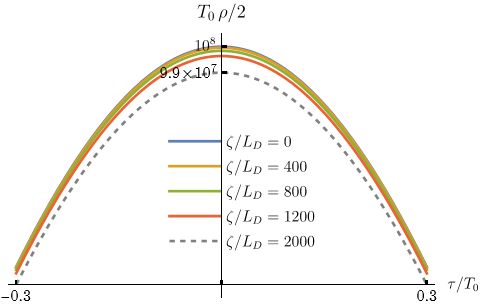}
         \caption{$-0.3<\tau/T_0<0.3$}
         \label{fig:ntotal}
     \end{subfigure}
     \hfill
     \begin{subfigure}[hb]{0.3\textwidth}
         \centering
         \includegraphics[width=\textwidth]{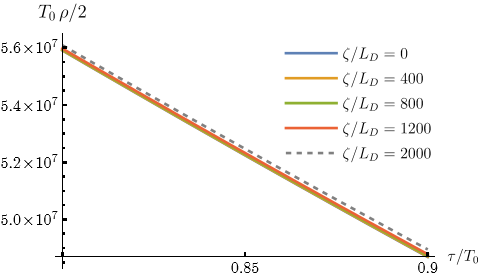}
         \caption{$1<\tau/T_0<1.1$}
         \label{fig:ntotal1}
     \end{subfigure}
     \hfill
     \begin{subfigure}[hb]{0.3\textwidth}
         \centering
         \includegraphics[width=\textwidth]{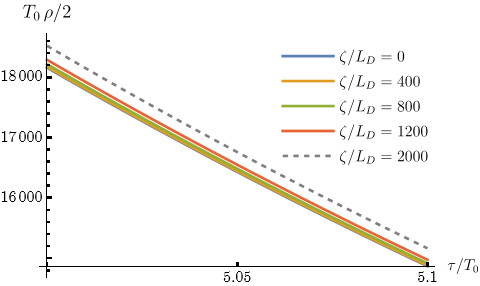}
         \caption{$5<\tau/T_0<5.1$}
         \label{fig:ntotal5}
     \end{subfigure}
        \caption{Total pulse photon number density. As the soliton propagates, the photon number density near the peak is reduced and moved to the front and back, symmetrically. Simulation for an initial pulse containing $10^8$ photons.}
        \label{fig:figmain3}
\end{figure}

In Fig.~\ref{fig:figmain2}, we plot the number density distortion ${2T_0}(\rho_\Delta+\rho_\c)$ and its $\zeta^2$ contribution $T_0L_D^2\partial_t^2 (\rho_\Delta+\rho_\c)$, respectively,  at $\zeta/L_D=1, 5, 50$.
For these propagation lengths larger than 10 dispersion lengths, we see that the distortion is dominated by vacuum instability and its backreaction. We note that, as we will demonstrate further below, our analysis is valid even much beyond these propagation distances.
Therefore, although the solution overall is affected by the choice of initial condition, consideration of the $\zeta^2$-dependent term allows us to identify the vacuum noise and backreaction effect as a fingerprint in the long propagation distance limit.
We also show the total photon number density, including the soliton background, in Fig.~\ref{fig:figmain3}.
We zoom in on the peak of the soliton and observe a small reduction of the peak value as the soliton propagates.
Since the total photon number of the soliton is conserved, the density in the front and back of the soliton increases, broadening the soliton as a result (See Fig.~\ref{fig:ntotal1} and Fig.~\ref{fig:ntotal5}).
In an experiment, the backreaction could be observed by measuring the spectral density of the soliton. Without the backreaction effect, the field will keep its $\sech$ shape as the soliton propagates.
Therefore, the backreaction effect can be identified as a deviation from the $\sech^2$ shape in the spectrum.
The Fourier transformed field is:
\begin{equation}
    \mathcal{F}(E_\c)(\omega)= \frac{\sqrt{P_0}(1+\omega^2T_0^2)}{2n_0}\sqrt{\frac{\pi}{2}}\sech(\pi\omega T_0/2)\frac{\zeta^2}{L_D^2},
\end{equation}
with a marked deviation from the classical soliton spectrum.

We now calculate the total photon number by integrating over all space $\tau$.
Again, we determine the photon number of each field contribution, Bogoliubov field and correction field,  and for the continuous and discrete modes, respectively.
For the Bogoliubov field, we obtain the continuous mode contribution $n_{\Delta,\con}$ and the discrete mode contribution $n_{\Delta,\dis}$ as:
\begin{align}
    n_{\Delta,\con} &:= \int \rho_{\Delta,\con} \ud \tau = \frac{1}{3},\label{eq:photonnumberConBog}\\
    n_{\Delta,\dis} &:= \int \rho_{\Delta,\dis} \ud \tau = \left(\frac{\pi^2}{18}+\frac{4}{6}+\frac{2}{3}\frac{\zeta^2}{L_D^2}\right).\label{eq:photonnumberDisBog}
\end{align}
Since the Bogoliubov field describes the quantum fluctuation of the soliton background, the sum of Eq.~\eqref{eq:photonnumberConBog} and Eq.~\eqref{eq:photonnumberDisBog} describes the expectation value of the quantum fluctuation of the soliton photon number.
Now we turn to the photon number from the backreaction effect.
The photon numbers derived from the existence of  the classical field correction from the continuous mode, $n_{\c,\con}$, and  from the discrete mode, $n_{\c,\dis}$, are, respectively: 
\begin{align}
    n_{\c,\con} &:=\int \rho_{\c,\con} \ud \tau = -\frac{2}{3},\label{eq:photonnumberConCor}\\
    n_{\c,\dis} &:= \int \rho_{\c,\dis} \ud \tau = \left(-\frac{13\pi^2}{90}-2-\frac{2}{3}\frac{\zeta^2}{L_D^2}\right).\label{eq:photonnumberDisCor}
\end{align}
Regarding backreaction effects, the important term is the $\zeta$-dependent term in the discrete mode in \eqref{eq:photonnumberDisCor} which describes the escape of photons from the soliton background.
Note that the $\zeta$-dependent terms are equal and opposite for the Bogoliubov and correction photon numbers, respectively,  which is essential for the number conservation.
The photon-number conservation can be checked by direct summation:
\begin{equation}
    n_\Delta+n_\c:=n_{\Delta,\con}+n_{\Delta,\dis}+n_{\c,\con}+n_{\c,\dis} = \mathrm{const.}
\end{equation}
Depending on the initial condition,  the total number of particles from the correction field ($ n_{\c,\con}$ and $n_{\c,\dis}$) may vary to give a different constant; only the $\zeta$-dependent term is not affected.

Finally, let us check the validity of approximations in our analysis.
Our assumptions are based on relation~\eqref{eq:ExpansionParameter}.
In our case, the assumption is valid if the Bogoliubov particle number is much smaller than the total particle number
\begin{equation}
    \frac{n_\Delta}{n_0}=\epsilon^2\xrightarrow[]{\zeta\gg L_D}\frac{2}{3n_0}\frac{\zeta^2}{L_D^2}\ll1.
\end{equation}
where $n_0$ is the total photon number.
For example, if the photon number of the initial pulse is $\sim10^8$ which is the typical experimental setup, our analysis holds up to about 1\% error ($\epsilon^2=0.01$) until $\zeta\sim 1.2\times10^3 L_D$, because for large $\zeta$,
\begin{equation}
    \frac{\zeta^2}{L_D^2} \simeq \frac{3 n_0}{2}\epsilon^2 =1.5\times10^{6}.
\end{equation}
\section{Conclusion}
We presented a complete nonrelativistic quantum field theoretic description of the fibre-optical soliton, the linear quantum fluctuation, and a backreaction effect in quadratic order.
Starting from the Lagrangian we developed a field expansion in order of the photon number that preserves the photon number. We quantised the field and analytically derived a complete set of modes, which consists of continuous and discrete modes. The discrete modes were shown to exhibit instabilities that characterise the field for propagation distances larger than tens of dispersion lengths. The discrete modes do not describe quasi-particle creation and annihilation, but instead are related to collective perturbations of the soliton parameters. 
Our result links in with the decay of solitons shown in \cite{Ward2023}, which is hence due to unstable discrete modes.
We show that because of the dominant behavior of the discrete modes, the photon density distortion develops a characteristic `crater' shape. Moreover, we suggest a measurement of the frequency distribution of the electric correction field as a signature of a backreaction effect.

The analogue backreaction research is one of the ways to approach quantum gravity with fluid \cite{Braunstein2023}.
The backreaction in leading order is a classical correction to the standard soliton solution.
This was calculated analytically for arbitrary initial conditions. As a result, the soliton temporally broadens and its peak value decreases quadratically in time. The deviations from the soliton take on a characteristic `crater' shape. It is straightforward to extend this analysis to an external dynamical source term acting on the soliton. 

Therefore, solitons are an attractive system for the observation of backreaction effects. Other systems considered in the past are BECs, quantum liquids and water wave \cite{Fischer2005,Fischer2007,Balbinot2005PRD,Balbinot2005PRL,Fagnocchi2006,Schuetzhold2008JPA,Schuetzhold2008PoS,Kurita2010,Tricella2020,Patrick2021,Baak2022,Braunstein2023}. In particular, these systems are seen as analogue gravity systems and can be considered test beds for the observation backreaction effects in the investigation of quantum fields in curved spacetime. 

\section{Acknowledgements}
We acknowledge helpful discussions with Natalia Korolkova, Sam Patrick and Christopher Burgess. 
This work was supported in part by the Science and Technology Facilities Council through the UKRI Quantum Technologies for Fundamental Physics Program [Grant ST/T005866/1 (FK)].
SB was supported by the National Research Foundation of Korea [Grant RS-2023-00247194].
\clearpage
\bibliography{SolQDyn}

\begin{thebibliography}{76}%
\makeatletter
\providecommand \@ifxundefined [1]{%
 \@ifx{#1\undefined}
}%
\providecommand \@ifnum [1]{%
 \ifnum #1\expandafter \@firstoftwo
 \else \expandafter \@secondoftwo
 \fi
}%
\providecommand \@ifx [1]{%
 \ifx #1\expandafter \@firstoftwo
 \else \expandafter \@secondoftwo
 \fi
}%
\providecommand \natexlab [1]{#1}%
\providecommand \enquote  [1]{``#1''}%
\providecommand \bibnamefont  [1]{#1}%
\providecommand \bibfnamefont [1]{#1}%
\providecommand \citenamefont [1]{#1}%
\providecommand \href@noop [0]{\@secondoftwo}%
\providecommand \href [0]{\begingroup \@sanitize@url \@href}%
\providecommand \@href[1]{\@@startlink{#1}\@@href}%
\providecommand \@@href[1]{\endgroup#1\@@endlink}%
\providecommand \@sanitize@url [0]{\catcode `\\12\catcode `\$12\catcode
  `\&12\catcode `\#12\catcode `\^12\catcode `\_12\catcode `\%12\relax}%
\providecommand \@@startlink[1]{}%
\providecommand \@@endlink[0]{}%
\providecommand \url  [0]{\begingroup\@sanitize@url \@url }%
\providecommand \@url [1]{\endgroup\@href {#1}{\urlprefix }}%
\providecommand \urlprefix  [0]{URL }%
\providecommand \Eprint [0]{\href }%
\providecommand \doibase [0]{https://doi.org/}%
\providecommand \selectlanguage [0]{\@gobble}%
\providecommand \bibinfo  [0]{\@secondoftwo}%
\providecommand \bibfield  [0]{\@secondoftwo}%
\providecommand \translation [1]{[#1]}%
\providecommand \BibitemOpen [0]{}%
\providecommand \bibitemStop [0]{}%
\providecommand \bibitemNoStop [0]{.\EOS\space}%
\providecommand \EOS [0]{\spacefactor3000\relax}%
\providecommand \BibitemShut  [1]{\csname bibitem#1\endcsname}%
\let\auto@bib@innerbib\@empty
\bibitem [{\citenamefont {Hollands}\ and\ \citenamefont
  {Wald}(2015)}]{Hollands2015}%
  \BibitemOpen
  \bibfield  {author} {\bibinfo {author} {\bibfnamefont {S.}~\bibnamefont
  {Hollands}}\ and\ \bibinfo {author} {\bibfnamefont {R.~M.}\ \bibnamefont
  {Wald}},\ }\bibfield  {title} {\bibinfo {title} {Quantum fields in curved
  spacetime},\ }\href
  {https://doi.org/https://doi.org/10.1016/j.physrep.2015.02.001} {\bibfield
  {journal} {\bibinfo  {journal} {Physics Reports}\ }\textbf {\bibinfo {volume}
  {574}},\ \bibinfo {pages} {1} (\bibinfo {year} {2015})},\ \bibinfo {note}
  {quantum fields in curved spacetime}\BibitemShut {NoStop}%
\bibitem [{\citenamefont {Fulling}(1989)}]{Fulling1989}%
  \BibitemOpen
  \bibfield  {author} {\bibinfo {author} {\bibfnamefont {S.~A.}\ \bibnamefont
  {Fulling}},\ }\href {https://doi.org/10.1017/CBO9781139172073} {\emph
  {\bibinfo {title} {{Aspects of Quantum Field Theory in Curved Spacetime}}}},\
  London Mathematical Society Student Texts\ (\bibinfo  {publisher} {Cambridge
  University Press},\ \bibinfo {year} {1989})\BibitemShut {NoStop}%
\bibitem [{\citenamefont {Kay}(2023)}]{Kay2023}%
  \BibitemOpen
  \bibfield  {author} {\bibinfo {author} {\bibfnamefont {B.~S.}\ \bibnamefont
  {Kay}},\ }\href@noop {} {\bibinfo {title} {{Quantum Field Theory in Curved
  Spacetime (2nd Edition)}}} (\bibinfo {year} {2023}),\ \Eprint
  {https://arxiv.org/abs/2308.14517} {arXiv:2308.14517 [gr-qc]} \BibitemShut
  {NoStop}%
\bibitem [{\citenamefont {Wald}(1994)}]{Wald1994}%
  \BibitemOpen
  \bibfield  {author} {\bibinfo {author} {\bibfnamefont {R.}~\bibnamefont
  {Wald}},\ }\href
  {https://press.uchicago.edu/ucp/books/book/chicago/Q/bo3684008.html} {\emph
  {\bibinfo {title} {{Quantum Field Theory in Curved Spacetime and Black Hole
  Thermodynamics}}}},\ Chicago Lectures in Physics\ (\bibinfo  {publisher}
  {University of Chicago Press},\ \bibinfo {year} {1994})\BibitemShut {NoStop}%
\bibitem [{\citenamefont {Mukhanov}\ and\ \citenamefont
  {Winitzki}(2007)}]{Mukhanov2007}%
  \BibitemOpen
  \bibfield  {author} {\bibinfo {author} {\bibfnamefont {V.}~\bibnamefont
  {Mukhanov}}\ and\ \bibinfo {author} {\bibfnamefont {S.}~\bibnamefont
  {Winitzki}},\ }\href {https://doi.org/10.1017/CBO9780511809149} {\emph
  {\bibinfo {title} {{Introduction to Quantum Effects in Gravity}}}}\ (\bibinfo
   {publisher} {Cambridge University Press},\ \bibinfo {year}
  {2007})\BibitemShut {NoStop}%
\bibitem [{\citenamefont {Ford}(1997)}]{Ford1997}%
  \BibitemOpen
  \bibfield  {author} {\bibinfo {author} {\bibfnamefont {L.~H.}\ \bibnamefont
  {Ford}},\ }\href {https://doi.org/10.48550/ARXIV.GR-QC/9707062} {\bibinfo
  {title} {{Quantum Field Theory in Curved Spacetime}}} (\bibinfo {year}
  {1997})\BibitemShut {NoStop}%
\bibitem [{\citenamefont {Parker}\ and\ \citenamefont
  {Toms}(2009)}]{Parker2009}%
  \BibitemOpen
  \bibfield  {author} {\bibinfo {author} {\bibfnamefont {L.}~\bibnamefont
  {Parker}}\ and\ \bibinfo {author} {\bibfnamefont {D.}~\bibnamefont {Toms}},\
  }\href@noop {} {\emph {\bibinfo {title} {{Quantum Field Theory in Curved
  Spacetime: Quantized Fields and Gravity}}}},\ Cambridge Monographs on
  Mathematical Physics\ (\bibinfo  {publisher} {Cambridge University Press},\
  \bibinfo {year} {2009})\BibitemShut {NoStop}%
\bibitem [{\citenamefont {Birell}\ and\ \citenamefont
  {Davies}(1982)}]{Birell1982}%
  \BibitemOpen
  \bibfield  {author} {\bibinfo {author} {\bibfnamefont {N.~D.}\ \bibnamefont
  {Birell}}\ and\ \bibinfo {author} {\bibfnamefont {P.~C.~W.}\ \bibnamefont
  {Davies}},\ }\href@noop {} {\emph {\bibinfo {title} {{Quantum Fields in
  Curved Space}}}}\ (\bibinfo  {publisher} {Cambridge University Press},\
  \bibinfo {year} {1982})\BibitemShut {NoStop}%
\bibitem [{\citenamefont {Wald}(1984)}]{Wald1984}%
  \BibitemOpen
  \bibfield  {author} {\bibinfo {author} {\bibfnamefont {R.~M.}\ \bibnamefont
  {Wald}},\ }\href
  {https://press.uchicago.edu/ucp/books/book/chicago/G/bo5952261.html} {\emph
  {\bibinfo {title} {{General relativity}}}}\ (\bibinfo  {publisher} {Chicago
  Univ. Press},\ \bibinfo {address} {Chicago, IL},\ \bibinfo {year}
  {1984})\BibitemShut {NoStop}%
\bibitem [{\citenamefont {Hawking}\ and\ \citenamefont
  {Ellis}(1973)}]{Hawking1973}%
  \BibitemOpen
  \bibfield  {author} {\bibinfo {author} {\bibfnamefont {S.~W.}\ \bibnamefont
  {Hawking}}\ and\ \bibinfo {author} {\bibfnamefont {G.~F.~R.}\ \bibnamefont
  {Ellis}},\ }\href {https://doi.org/10.1017/CBO9780511524646} {\emph {\bibinfo
  {title} {{The Large Scale Structure of Space-Time}}}},\ Cambridge Monographs
  on Mathematical Physics\ (\bibinfo  {publisher} {Cambridge University
  Press},\ \bibinfo {year} {1973})\BibitemShut {NoStop}%
\bibitem [{\citenamefont {Misner}\ \emph {et~al.}(2017)\citenamefont {Misner},
  \citenamefont {Thorne}, \citenamefont {Wheeler},\ and\ \citenamefont
  {Kaiser}}]{MTW2017}%
  \BibitemOpen
  \bibfield  {author} {\bibinfo {author} {\bibfnamefont {C.}~\bibnamefont
  {Misner}}, \bibinfo {author} {\bibfnamefont {K.}~\bibnamefont {Thorne}},
  \bibinfo {author} {\bibfnamefont {J.}~\bibnamefont {Wheeler}},\ and\ \bibinfo
  {author} {\bibfnamefont {D.}~\bibnamefont {Kaiser}},\ }\href
  {https://press.princeton.edu/books/hardcover/9780691177793/gravitation}
  {\emph {\bibinfo {title} {{Gravitation}}}}\ (\bibinfo  {publisher} {Princeton
  University Press},\ \bibinfo {year} {2017})\BibitemShut {NoStop}%
\bibitem [{\citenamefont {Schander}\ and\ \citenamefont
  {Thiemann}(2021)}]{Schander2021}%
  \BibitemOpen
  \bibfield  {author} {\bibinfo {author} {\bibfnamefont {S.}~\bibnamefont
  {Schander}}\ and\ \bibinfo {author} {\bibfnamefont {T.}~\bibnamefont
  {Thiemann}},\ }\bibfield  {title} {\bibinfo {title} {{Backreaction in
  Cosmology}},\ }\bibfield  {journal} {\bibinfo  {journal} {Frontiers in
  Astronomy and Space Sciences}\ }\textbf {\bibinfo {volume} {8}},\ \href
  {https://doi.org/10.3389/fspas.2021.692198} {10.3389/fspas.2021.692198}
  (\bibinfo {year} {2021})\BibitemShut {NoStop}%
\bibitem [{\citenamefont {Unruh}(1981)}]{Unruh1981}%
  \BibitemOpen
  \bibfield  {author} {\bibinfo {author} {\bibfnamefont {W.~G.}\ \bibnamefont
  {Unruh}},\ }\bibfield  {title} {\bibinfo {title} {{Experimental Black-Hole
  Evaporation?}},\ }\href {https://doi.org/10.1103/PhysRevLett.46.1351}
  {\bibfield  {journal} {\bibinfo  {journal} {Phys. Rev. Lett.}\ }\textbf
  {\bibinfo {volume} {46}},\ \bibinfo {pages} {1351} (\bibinfo {year}
  {1981})}\BibitemShut {NoStop}%
\bibitem [{\citenamefont {Philbin}\ \emph {et~al.}(2008)\citenamefont
  {Philbin}, \citenamefont {Kuklewicz}, \citenamefont {Robertson},
  \citenamefont {Hill}, \citenamefont {König},\ and\ \citenamefont
  {Leonhardt}}]{Philbin2008}%
  \BibitemOpen
  \bibfield  {author} {\bibinfo {author} {\bibfnamefont {T.~G.}\ \bibnamefont
  {Philbin}}, \bibinfo {author} {\bibfnamefont {C.}~\bibnamefont {Kuklewicz}},
  \bibinfo {author} {\bibfnamefont {S.}~\bibnamefont {Robertson}}, \bibinfo
  {author} {\bibfnamefont {S.}~\bibnamefont {Hill}}, \bibinfo {author}
  {\bibfnamefont {F.}~\bibnamefont {König}},\ and\ \bibinfo {author}
  {\bibfnamefont {U.}~\bibnamefont {Leonhardt}},\ }\bibfield  {title} {\bibinfo
  {title} {{Fiber-Optical Analog of the Event Horizon}},\ }\href
  {https://doi.org/10.1126/science.1153625} {\bibfield  {journal} {\bibinfo
  {journal} {Science}\ }\textbf {\bibinfo {volume} {319}},\ \bibinfo {pages}
  {1367} (\bibinfo {year} {2008})},\ \Eprint
  {https://arxiv.org/abs/https://www.science.org/doi/pdf/10.1126/science.1153625}
  {https://www.science.org/doi/pdf/10.1126/science.1153625} \BibitemShut
  {NoStop}%
\bibitem [{\citenamefont {Weinfurtner}\ \emph {et~al.}(2011)\citenamefont
  {Weinfurtner}, \citenamefont {Tedford}, \citenamefont {Penrice},
  \citenamefont {Unruh},\ and\ \citenamefont {Lawrence}}]{Silke2011}%
  \BibitemOpen
  \bibfield  {author} {\bibinfo {author} {\bibfnamefont {S.}~\bibnamefont
  {Weinfurtner}}, \bibinfo {author} {\bibfnamefont {E.~W.}\ \bibnamefont
  {Tedford}}, \bibinfo {author} {\bibfnamefont {M.~C.~J.}\ \bibnamefont
  {Penrice}}, \bibinfo {author} {\bibfnamefont {W.~G.}\ \bibnamefont {Unruh}},\
  and\ \bibinfo {author} {\bibfnamefont {G.~A.}\ \bibnamefont {Lawrence}},\
  }\bibfield  {title} {\bibinfo {title} {{Measurement of Stimulated Hawking
  Emission in an Analogue System}},\ }\href
  {https://doi.org/10.1103/PhysRevLett.106.021302} {\bibfield  {journal}
  {\bibinfo  {journal} {Phys. Rev. Lett.}\ }\textbf {\bibinfo {volume} {106}},\
  \bibinfo {pages} {021302} (\bibinfo {year} {2011})}\BibitemShut {NoStop}%
\bibitem [{\citenamefont {Steinhauer}(2016)}]{Steinhauer2016}%
  \BibitemOpen
  \bibfield  {author} {\bibinfo {author} {\bibfnamefont {J.}~\bibnamefont
  {Steinhauer}},\ }\bibfield  {title} {\bibinfo {title} {{Observation of
  quantum Hawking radiation and its entanglement in an analogue black hole}},\
  }\href {http://dx.doi.org/10.1038/nphys3863} {\bibfield  {journal} {\bibinfo
  {journal} {Nat. Phys.}\ }\textbf {\bibinfo {volume} {12}},\ \bibinfo {pages}
  {959} (\bibinfo {year} {2016})}\BibitemShut {NoStop}%
\bibitem [{\citenamefont {Mu{\~n}oz~de Nova}\ \emph {et~al.}(2019)\citenamefont
  {Mu{\~n}oz~de Nova}, \citenamefont {Golubkov}, \citenamefont {Kolobov},\ and\
  \citenamefont {Steinhauer}}]{Steinhauer2019}%
  \BibitemOpen
  \bibfield  {author} {\bibinfo {author} {\bibfnamefont {J.~R.}\ \bibnamefont
  {Mu{\~n}oz~de Nova}}, \bibinfo {author} {\bibfnamefont {K.}~\bibnamefont
  {Golubkov}}, \bibinfo {author} {\bibfnamefont {V.~I.}\ \bibnamefont
  {Kolobov}},\ and\ \bibinfo {author} {\bibfnamefont {J.}~\bibnamefont
  {Steinhauer}},\ }\bibfield  {title} {\bibinfo {title} {{Observation of
  thermal Hawking radiation and its temperature in an analogue black hole}},\
  }\href {https://doi.org/10.1038/s41586-019-1241-0} {\bibfield  {journal}
  {\bibinfo  {journal} {Nature}\ }\textbf {\bibinfo {volume} {569}},\ \bibinfo
  {pages} {688} (\bibinfo {year} {2019})}\BibitemShut {NoStop}%
\bibitem [{\citenamefont {Sch\"utzhold}\ \emph {et~al.}(2005)\citenamefont
  {Sch\"utzhold}, \citenamefont {Uhlmann}, \citenamefont {Xu},\ and\
  \citenamefont {Fischer}}]{Fischer2005}%
  \BibitemOpen
  \bibfield  {author} {\bibinfo {author} {\bibfnamefont {R.}~\bibnamefont
  {Sch\"utzhold}}, \bibinfo {author} {\bibfnamefont {M.}~\bibnamefont
  {Uhlmann}}, \bibinfo {author} {\bibfnamefont {Y.}~\bibnamefont {Xu}},\ and\
  \bibinfo {author} {\bibfnamefont {U.~R.}\ \bibnamefont {Fischer}},\
  }\bibfield  {title} {\bibinfo {title} {{Quantum backreaction in dilute
  Bose-Einstein condensates}},\ }\href
  {https://doi.org/10.1103/PhysRevD.72.105005} {\bibfield  {journal} {\bibinfo
  {journal} {Phys. Rev. D}\ }\textbf {\bibinfo {volume} {72}},\ \bibinfo
  {pages} {105005} (\bibinfo {year} {2005})}\BibitemShut {NoStop}%
\bibitem [{\citenamefont {Fischer}(2007)}]{Fischer2007}%
  \BibitemOpen
  \bibfield  {author} {\bibinfo {author} {\bibfnamefont {U.~R.}\ \bibnamefont
  {Fischer}},\ }\bibinfo {title} {{Dynamical Aspects of Analogue Gravity: The
  Backreaction of Quantum Fluctuations in Dilute Bose-Einstein Condensates}},\
  in\ \href {https://doi.org/10.1007/3-540-70859-6_5} {\emph {\bibinfo
  {booktitle} {{Quantum Analogues: From Phase Transitions to Black Holes and
  Cosmology}}}},\ \bibinfo {editor} {edited by\ \bibinfo {editor}
  {\bibfnamefont {W.~G.}\ \bibnamefont {Unruh}}\ and\ \bibinfo {editor}
  {\bibfnamefont {R.}~\bibnamefont {Sch{\"u}tzhold}}}\ (\bibinfo  {publisher}
  {Springer Berlin Heidelberg},\ \bibinfo {address} {Berlin, Heidelberg},\
  \bibinfo {year} {2007})\ pp.\ \bibinfo {pages} {93--113}\BibitemShut
  {NoStop}%
\bibitem [{\citenamefont {Balbinot}\ \emph
  {et~al.}(2005{\natexlab{a}})\citenamefont {Balbinot}, \citenamefont
  {Fagnocchi},\ and\ \citenamefont {Fabbri}}]{Balbinot2005PRD}%
  \BibitemOpen
  \bibfield  {author} {\bibinfo {author} {\bibfnamefont {R.}~\bibnamefont
  {Balbinot}}, \bibinfo {author} {\bibfnamefont {S.}~\bibnamefont
  {Fagnocchi}},\ and\ \bibinfo {author} {\bibfnamefont {A.}~\bibnamefont
  {Fabbri}},\ }\bibfield  {title} {\bibinfo {title} {{Quantum effects in
  acoustic black holes: The backreaction}},\ }\href
  {https://doi.org/10.1103/PhysRevD.71.064019} {\bibfield  {journal} {\bibinfo
  {journal} {Phys. Rev. D}\ }\textbf {\bibinfo {volume} {71}},\ \bibinfo
  {pages} {064019} (\bibinfo {year} {2005}{\natexlab{a}})}\BibitemShut
  {NoStop}%
\bibitem [{\citenamefont {Balbinot}\ \emph
  {et~al.}(2005{\natexlab{b}})\citenamefont {Balbinot}, \citenamefont
  {Fagnocchi}, \citenamefont {Fabbri},\ and\ \citenamefont
  {Procopio}}]{Balbinot2005PRL}%
  \BibitemOpen
  \bibfield  {author} {\bibinfo {author} {\bibfnamefont {R.}~\bibnamefont
  {Balbinot}}, \bibinfo {author} {\bibfnamefont {S.}~\bibnamefont {Fagnocchi}},
  \bibinfo {author} {\bibfnamefont {A.}~\bibnamefont {Fabbri}},\ and\ \bibinfo
  {author} {\bibfnamefont {G.~P.}\ \bibnamefont {Procopio}},\ }\bibfield
  {title} {\bibinfo {title} {{Backreaction in Acoustic Black Holes}},\ }\href
  {https://doi.org/10.1103/PhysRevLett.94.161302} {\bibfield  {journal}
  {\bibinfo  {journal} {Phys. Rev. Lett.}\ }\textbf {\bibinfo {volume} {94}},\
  \bibinfo {pages} {161302} (\bibinfo {year} {2005}{\natexlab{b}})}\BibitemShut
  {NoStop}%
\bibitem [{\citenamefont {Fagnocchi}(2006)}]{Fagnocchi2006}%
  \BibitemOpen
  \bibfield  {author} {\bibinfo {author} {\bibfnamefont {S.}~\bibnamefont
  {Fagnocchi}},\ }\bibfield  {title} {\bibinfo {title} {{Back-reaction in
  acoustic black holes}},\ }\href {https://doi.org/10.1088/1742-6596/33/1/057}
  {\bibfield  {journal} {\bibinfo  {journal} {Journal of Physics: Conference
  Series}\ }\textbf {\bibinfo {volume} {33}},\ \bibinfo {pages} {445} (\bibinfo
  {year} {2006})}\BibitemShut {NoStop}%
\bibitem [{\citenamefont {Schützhold}\ and\ \citenamefont
  {Maia}(2008)}]{Schuetzhold2008JPA}%
  \BibitemOpen
  \bibfield  {author} {\bibinfo {author} {\bibfnamefont {R.}~\bibnamefont
  {Schützhold}}\ and\ \bibinfo {author} {\bibfnamefont {C.}~\bibnamefont
  {Maia}},\ }\bibfield  {title} {\bibinfo {title} {{Black-hole
  back-reaction—a toy model}},\ }\href
  {https://doi.org/10.1088/1751-8113/41/16/164065} {\bibfield  {journal}
  {\bibinfo  {journal} {Journal of Physics A: Mathematical and Theoretical}\
  }\textbf {\bibinfo {volume} {41}},\ \bibinfo {pages} {164065} (\bibinfo
  {year} {2008})}\BibitemShut {NoStop}%
\bibitem [{\citenamefont {Sch{\"u}tzhold}(2008)}]{Schuetzhold2008PoS}%
  \BibitemOpen
  \bibfield  {author} {\bibinfo {author} {\bibfnamefont {R.}~\bibnamefont
  {Sch{\"u}tzhold}},\ }\bibfield  {title} {\bibinfo {title} {{Quantum
  back-reaction problems}},\ }in\ \href {https://doi.org/10.22323/1.043.0036}
  {\emph {\bibinfo {booktitle} {Proceedings of From Quantum to Emergent
  Gravity: Theory and Phenomenology {\textemdash} PoS(QG-Ph)}}},\ Vol.\
  \bibinfo {volume} {043}\ (\bibinfo {year} {2008})\ p.\ \bibinfo {pages}
  {036}\BibitemShut {NoStop}%
\bibitem [{\citenamefont {Kurita}\ \emph {et~al.}(2010)\citenamefont {Kurita},
  \citenamefont {Kobayashi}, \citenamefont {Ishihara},\ and\ \citenamefont
  {Tsubota}}]{Kurita2010}%
  \BibitemOpen
  \bibfield  {author} {\bibinfo {author} {\bibfnamefont {Y.}~\bibnamefont
  {Kurita}}, \bibinfo {author} {\bibfnamefont {M.}~\bibnamefont {Kobayashi}},
  \bibinfo {author} {\bibfnamefont {H.}~\bibnamefont {Ishihara}},\ and\
  \bibinfo {author} {\bibfnamefont {M.}~\bibnamefont {Tsubota}},\ }\bibfield
  {title} {\bibinfo {title} {{Particle creation in Bose-Einstein condensates:
  Theoretical formulation based on conserving gapless mean-field theory}},\
  }\href {https://doi.org/10.1103/PhysRevA.82.053602} {\bibfield  {journal}
  {\bibinfo  {journal} {Phys. Rev. A}\ }\textbf {\bibinfo {volume} {82}},\
  \bibinfo {pages} {053602} (\bibinfo {year} {2010})}\BibitemShut {NoStop}%
\bibitem [{\citenamefont {Liberati}\ \emph {et~al.}(2020)\citenamefont
  {Liberati}, \citenamefont {Tricella},\ and\ \citenamefont
  {Trombettoni}}]{Tricella2020}%
  \BibitemOpen
  \bibfield  {author} {\bibinfo {author} {\bibfnamefont {S.}~\bibnamefont
  {Liberati}}, \bibinfo {author} {\bibfnamefont {G.}~\bibnamefont {Tricella}},\
  and\ \bibinfo {author} {\bibfnamefont {A.}~\bibnamefont {Trombettoni}},\
  }\bibfield  {title} {\bibinfo {title} {{Back-Reaction in Canonical Analogue
  Black Holes}},\ }\bibfield  {journal} {\bibinfo  {journal} {Applied
  Sciences}\ }\textbf {\bibinfo {volume} {10}},\ \href
  {https://doi.org/10.3390/app10248868} {10.3390/app10248868} (\bibinfo {year}
  {2020})\BibitemShut {NoStop}%
\bibitem [{\citenamefont {Patrick}\ \emph {et~al.}(2021)\citenamefont
  {Patrick}, \citenamefont {Goodhew}, \citenamefont {Gooding},\ and\
  \citenamefont {Weinfurtner}}]{Patrick2021}%
  \BibitemOpen
  \bibfield  {author} {\bibinfo {author} {\bibfnamefont {S.}~\bibnamefont
  {Patrick}}, \bibinfo {author} {\bibfnamefont {H.}~\bibnamefont {Goodhew}},
  \bibinfo {author} {\bibfnamefont {C.}~\bibnamefont {Gooding}},\ and\ \bibinfo
  {author} {\bibfnamefont {S.}~\bibnamefont {Weinfurtner}},\ }\bibfield
  {title} {\bibinfo {title} {{Backreaction in an Analogue Black Hole
  Experimen}t},\ }\href {https://doi.org/10.1103/PhysRevLett.126.041105}
  {\bibfield  {journal} {\bibinfo  {journal} {Phys. Rev. Lett.}\ }\textbf
  {\bibinfo {volume} {126}},\ \bibinfo {pages} {041105} (\bibinfo {year}
  {2021})}\BibitemShut {NoStop}%
\bibitem [{\citenamefont {Baak}\ \emph {et~al.}(2022)\citenamefont {Baak},
  \citenamefont {Ribeiro},\ and\ \citenamefont {Fischer}}]{Baak2022}%
  \BibitemOpen
  \bibfield  {author} {\bibinfo {author} {\bibfnamefont {S.-S.}\ \bibnamefont
  {Baak}}, \bibinfo {author} {\bibfnamefont {C.~C.~H.}\ \bibnamefont
  {Ribeiro}},\ and\ \bibinfo {author} {\bibfnamefont {U.~R.}\ \bibnamefont
  {Fischer}},\ }\bibfield  {title} {\bibinfo {title} {{Number-conserving
  solution for dynamical quantum backreaction in a Bose-Einstein condensate}},\
  }\href {https://doi.org/10.1103/PhysRevA.106.053319} {\bibfield  {journal}
  {\bibinfo  {journal} {Phys. Rev. A}\ }\textbf {\bibinfo {volume} {106}},\
  \bibinfo {pages} {053319} (\bibinfo {year} {2022})}\BibitemShut {NoStop}%
\bibitem [{\citenamefont {Braunstein}\ \emph {et~al.}(2023)\citenamefont
  {Braunstein}, \citenamefont {Faizal}, \citenamefont {Krauss}, \citenamefont
  {Marino},\ and\ \citenamefont {Shah}}]{Braunstein2023}%
  \BibitemOpen
  \bibfield  {author} {\bibinfo {author} {\bibfnamefont {S.~L.}\ \bibnamefont
  {Braunstein}}, \bibinfo {author} {\bibfnamefont {M.}~\bibnamefont {Faizal}},
  \bibinfo {author} {\bibfnamefont {L.~M.}\ \bibnamefont {Krauss}}, \bibinfo
  {author} {\bibfnamefont {F.}~\bibnamefont {Marino}},\ and\ \bibinfo {author}
  {\bibfnamefont {N.~A.}\ \bibnamefont {Shah}},\ }\bibfield  {title} {\bibinfo
  {title} {Analogue simulations of quantum gravity with fluids},\ }\href
  {https://doi.org/10.1038/s42254-023-00630-y} {\bibfield  {journal} {\bibinfo
  {journal} {Nature Reviews Physics}\ }\textbf {\bibinfo {volume} {5}},\
  \bibinfo {pages} {612} (\bibinfo {year} {2023})}\BibitemShut {NoStop}%
\bibitem [{\citenamefont {Choudhary}\ and\ \citenamefont
  {K\"{o}nig}(2012)}]{Choudhary2012}%
  \BibitemOpen
  \bibfield  {author} {\bibinfo {author} {\bibfnamefont {A.}~\bibnamefont
  {Choudhary}}\ and\ \bibinfo {author} {\bibfnamefont {F.}~\bibnamefont
  {K\"{o}nig}},\ }\bibfield  {title} {\bibinfo {title} {Efficient frequency
  shifting of dispersive waves at solitons},\ }\href
  {https://doi.org/10.1364/OE.20.005538} {\bibfield  {journal} {\bibinfo
  {journal} {Opt. Express}\ }\textbf {\bibinfo {volume} {20}},\ \bibinfo
  {pages} {5538} (\bibinfo {year} {2012})}\BibitemShut {NoStop}%
\bibitem [{\citenamefont {Villari}\ \emph {et~al.}(2018)\citenamefont
  {Villari}, \citenamefont {Faccio}, \citenamefont {Biancalana},\ and\
  \citenamefont {Conti}}]{Villari2018}%
  \BibitemOpen
  \bibfield  {author} {\bibinfo {author} {\bibfnamefont {L.~D.~M.}\
  \bibnamefont {Villari}}, \bibinfo {author} {\bibfnamefont {D.}~\bibnamefont
  {Faccio}}, \bibinfo {author} {\bibfnamefont {F.}~\bibnamefont {Biancalana}},\
  and\ \bibinfo {author} {\bibfnamefont {C.}~\bibnamefont {Conti}},\ }\bibfield
   {title} {\bibinfo {title} {{Quantum soliton evaporation}},\ }\href
  {https://doi.org/10.1103/PhysRevA.98.043859} {\bibfield  {journal} {\bibinfo
  {journal} {Phys. Rev. A}\ }\textbf {\bibinfo {volume} {98}},\ \bibinfo
  {pages} {043859} (\bibinfo {year} {2018})}\BibitemShut {NoStop}%
\bibitem [{\citenamefont {Ward}\ \emph {et~al.}(2023)\citenamefont {Ward},
  \citenamefont {Allahverdi},\ and\ \citenamefont {Mafi}}]{Ward2023}%
  \BibitemOpen
  \bibfield  {author} {\bibinfo {author} {\bibfnamefont {S.}~\bibnamefont
  {Ward}}, \bibinfo {author} {\bibfnamefont {R.}~\bibnamefont {Allahverdi}},\
  and\ \bibinfo {author} {\bibfnamefont {A.}~\bibnamefont {Mafi}},\ }\bibfield
  {title} {\bibinfo {title} {{Quantum decay of an optical soliton}},\ }\href
  {https://doi.org/10.1103/PhysRevA.107.053513} {\bibfield  {journal} {\bibinfo
   {journal} {Phys. Rev. A}\ }\textbf {\bibinfo {volume} {107}},\ \bibinfo
  {pages} {053513} (\bibinfo {year} {2023})}\BibitemShut {NoStop}%
\bibitem [{\citenamefont {Girardeau}\ and\ \citenamefont
  {Arnowitt}(1959)}]{Girardeau1959}%
  \BibitemOpen
  \bibfield  {author} {\bibinfo {author} {\bibfnamefont {M.}~\bibnamefont
  {Girardeau}}\ and\ \bibinfo {author} {\bibfnamefont {R.}~\bibnamefont
  {Arnowitt}},\ }\bibfield  {title} {\bibinfo {title} {{Theory of Many-Boson
  Systems: Pair Theory}},\ }\href {https://doi.org/10.1103/PhysRev.113.755}
  {\bibfield  {journal} {\bibinfo  {journal} {Phys. Rev.}\ }\textbf {\bibinfo
  {volume} {113}},\ \bibinfo {pages} {755} (\bibinfo {year}
  {1959})}\BibitemShut {NoStop}%
\bibitem [{\citenamefont {Gardiner}(1997)}]{Gardiner1997}%
  \BibitemOpen
  \bibfield  {author} {\bibinfo {author} {\bibfnamefont {C.~W.}\ \bibnamefont
  {Gardiner}},\ }\bibfield  {title} {\bibinfo {title}
  {{Particle-number-conserving Bogoliubov method which demonstrates the
  validity of the time-dependent Gross-Pitaevskii equation for a highly
  condensed Bose gas}},\ }\href {https://doi.org/10.1103/PhysRevA.56.1414}
  {\bibfield  {journal} {\bibinfo  {journal} {Phys. Rev. A}\ }\textbf {\bibinfo
  {volume} {56}},\ \bibinfo {pages} {1414} (\bibinfo {year}
  {1997})}\BibitemShut {NoStop}%
\bibitem [{\citenamefont {Girardeau}(1998)}]{Girardeau1998}%
  \BibitemOpen
  \bibfield  {author} {\bibinfo {author} {\bibfnamefont {M.~D.}\ \bibnamefont
  {Girardeau}},\ }\bibfield  {title} {\bibinfo {title} {{Comment on
  ``Particle-number-conserving Bogoliubov method which demonstrates the
  validity of the time-dependent Gross-Pitaevskii equation for a highly
  condensed Bose gas''}},\ }\href {https://doi.org/10.1103/PhysRevA.58.775}
  {\bibfield  {journal} {\bibinfo  {journal} {Phys. Rev. A}\ }\textbf {\bibinfo
  {volume} {58}},\ \bibinfo {pages} {775} (\bibinfo {year} {1998})}\BibitemShut
  {NoStop}%
\bibitem [{\citenamefont {Castin}\ and\ \citenamefont
  {Dum}(1998)}]{Castin1998}%
  \BibitemOpen
  \bibfield  {author} {\bibinfo {author} {\bibfnamefont {Y.}~\bibnamefont
  {Castin}}\ and\ \bibinfo {author} {\bibfnamefont {R.}~\bibnamefont {Dum}},\
  }\bibfield  {title} {\bibinfo {title} {{Low-temperature Bose-Einstein
  condensates in time-dependent traps: Beyond the $U(1)$ symmetry-breaking
  approach}},\ }\href {https://doi.org/10.1103/PhysRevA.57.3008} {\bibfield
  {journal} {\bibinfo  {journal} {Phys. Rev. A}\ }\textbf {\bibinfo {volume}
  {57}},\ \bibinfo {pages} {3008} (\bibinfo {year} {1998})}\BibitemShut
  {NoStop}%
\bibitem [{\citenamefont {Lieb}\ \emph {et~al.}(2000)\citenamefont {Lieb},
  \citenamefont {Seiringer},\ and\ \citenamefont {Yngvason}}]{Lieb2000}%
  \BibitemOpen
  \bibfield  {author} {\bibinfo {author} {\bibfnamefont {E.~H.}\ \bibnamefont
  {Lieb}}, \bibinfo {author} {\bibfnamefont {R.}~\bibnamefont {Seiringer}},\
  and\ \bibinfo {author} {\bibfnamefont {J.}~\bibnamefont {Yngvason}},\
  }\bibfield  {title} {\bibinfo {title} {{Bosons in a trap: A rigorous
  derivation of the Gross-Pitaevskii energy functional}},\ }\href
  {https://doi.org/10.1103/PhysRevA.61.043602} {\bibfield  {journal} {\bibinfo
  {journal} {Phys. Rev. A}\ }\textbf {\bibinfo {volume} {61}},\ \bibinfo
  {pages} {043602} (\bibinfo {year} {2000})}\BibitemShut {NoStop}%
\bibitem [{\citenamefont {Gardiner}\ and\ \citenamefont
  {Morgan}(2007)}]{Gardiner2007}%
  \BibitemOpen
  \bibfield  {author} {\bibinfo {author} {\bibfnamefont {S.~A.}\ \bibnamefont
  {Gardiner}}\ and\ \bibinfo {author} {\bibfnamefont {S.~A.}\ \bibnamefont
  {Morgan}},\ }\bibfield  {title} {\bibinfo {title} {{Number-conserving
  approach to a minimal self-consistent treatment of condensate and
  noncondensate dynamics in a degenerate Bose gas}},\ }\href
  {https://doi.org/10.1103/PhysRevA.75.043621} {\bibfield  {journal} {\bibinfo
  {journal} {Phys. Rev. A}\ }\textbf {\bibinfo {volume} {75}},\ \bibinfo
  {pages} {043621} (\bibinfo {year} {2007})}\BibitemShut {NoStop}%
\bibitem [{\citenamefont {Billam}\ and\ \citenamefont
  {Gardiner}(2012)}]{Billam2012}%
  \BibitemOpen
  \bibfield  {author} {\bibinfo {author} {\bibfnamefont {T.~P.}\ \bibnamefont
  {Billam}}\ and\ \bibinfo {author} {\bibfnamefont {S.~A.}\ \bibnamefont
  {Gardiner}},\ }\bibfield  {title} {\bibinfo {title} {{Coherence and
  instability in a driven Bose{\textendash}Einstein condensate: a fully
  dynamical number-conserving approach}},\ }\href
  {https://doi.org/10.1088/1367-2630/14/1/013038} {\bibfield  {journal}
  {\bibinfo  {journal} {New Journal of Physics}\ }\textbf {\bibinfo {volume}
  {14}},\ \bibinfo {pages} {013038} (\bibinfo {year} {2012})}\BibitemShut
  {NoStop}%
\bibitem [{\citenamefont {Billam}\ \emph {et~al.}(2013)\citenamefont {Billam},
  \citenamefont {Mason},\ and\ \citenamefont {Gardiner}}]{Billam2013}%
  \BibitemOpen
  \bibfield  {author} {\bibinfo {author} {\bibfnamefont {T.~P.}\ \bibnamefont
  {Billam}}, \bibinfo {author} {\bibfnamefont {P.}~\bibnamefont {Mason}},\ and\
  \bibinfo {author} {\bibfnamefont {S.~A.}\ \bibnamefont {Gardiner}},\
  }\bibfield  {title} {\bibinfo {title} {{Second-order number-conserving
  description of nonequilibrium dynamics in finite-temperature Bose-Einstein
  condensates}},\ }\href {https://doi.org/10.1103/PhysRevA.87.033628}
  {\bibfield  {journal} {\bibinfo  {journal} {Phys. Rev. A}\ }\textbf {\bibinfo
  {volume} {87}},\ \bibinfo {pages} {033628} (\bibinfo {year}
  {2013})}\BibitemShut {NoStop}%
\bibitem [{\citenamefont {et~al.}(1989{\natexlab{a}})}]{Lai2}%
  \BibitemOpen
  \bibfield  {author} {\bibinfo {author} {\bibfnamefont {Y.~L.}\ \bibnamefont
  {et~al.}},\ }\bibfield  {title} {\bibinfo {title} {{Quantum theory of
  solitons in optical fibers. II. Exact solution}},\ }\href
  {https://doi.org/10.1103/PhysRevA.40.854} {\bibfield  {journal} {\bibinfo
  {journal} {Phys.Rev.A}\ }\textbf {\bibinfo {volume} {40}},\ \bibinfo {pages}
  {854} (\bibinfo {year} {1989}{\natexlab{a}})}\BibitemShut {NoStop}%
\bibitem [{\citenamefont {et~al.}(1989{\natexlab{b}})}]{Lai1}%
  \BibitemOpen
  \bibfield  {author} {\bibinfo {author} {\bibfnamefont {Y.~L.}\ \bibnamefont
  {et~al.}},\ }\bibfield  {title} {\bibinfo {title} {{Quantum theory of
  solitons in optical fibers. I. Time-dependent Hartree approximation}},\
  }\href {https://doi.org/10.1103/PhysRevA.40.844} {\bibfield  {journal}
  {\bibinfo  {journal} {Phys.Rev.A}\ }\textbf {\bibinfo {volume} {40}},\
  \bibinfo {pages} {844} (\bibinfo {year} {1989}{\natexlab{b}})}\BibitemShut
  {NoStop}%
\bibitem [{\citenamefont {Carter}\ \emph {et~al.}(1987)\citenamefont {Carter},
  \citenamefont {Drummond}, \citenamefont {Reid},\ and\ \citenamefont
  {Shelby}}]{Carter1987}%
  \BibitemOpen
  \bibfield  {author} {\bibinfo {author} {\bibfnamefont {S.~J.}\ \bibnamefont
  {Carter}}, \bibinfo {author} {\bibfnamefont {P.~D.}\ \bibnamefont
  {Drummond}}, \bibinfo {author} {\bibfnamefont {M.~D.}\ \bibnamefont {Reid}},\
  and\ \bibinfo {author} {\bibfnamefont {R.~M.}\ \bibnamefont {Shelby}},\
  }\bibfield  {title} {\bibinfo {title} {Squeezing of quantum solitons},\
  }\href {https://doi.org/10.1103/PhysRevLett.58.1841} {\bibfield  {journal}
  {\bibinfo  {journal} {Phys. Rev. Lett.}\ }\textbf {\bibinfo {volume} {58}},\
  \bibinfo {pages} {1841} (\bibinfo {year} {1987})}\BibitemShut {NoStop}%
\bibitem [{\citenamefont {Sp\"alter}\ \emph {et~al.}(1998)\citenamefont
  {Sp\"alter}, \citenamefont {Korolkova}, \citenamefont {K\"onig},
  \citenamefont {Sizmann},\ and\ \citenamefont {Leuchs}}]{Spalter1998PRL}%
  \BibitemOpen
  \bibfield  {author} {\bibinfo {author} {\bibfnamefont {S.}~\bibnamefont
  {Sp\"alter}}, \bibinfo {author} {\bibfnamefont {N.}~\bibnamefont
  {Korolkova}}, \bibinfo {author} {\bibfnamefont {F.}~\bibnamefont {K\"onig}},
  \bibinfo {author} {\bibfnamefont {A.}~\bibnamefont {Sizmann}},\ and\ \bibinfo
  {author} {\bibfnamefont {G.}~\bibnamefont {Leuchs}},\ }\bibfield  {title}
  {\bibinfo {title} {Observation of multimode quantum correlations in fiber
  optical solitons},\ }\href {https://doi.org/10.1103/PhysRevLett.81.786}
  {\bibfield  {journal} {\bibinfo  {journal} {Phys. Rev. Lett.}\ }\textbf
  {\bibinfo {volume} {81}},\ \bibinfo {pages} {786} (\bibinfo {year}
  {1998})}\BibitemShut {NoStop}%
\bibitem [{\citenamefont {Schmitt}\ \emph {et~al.}(1998)\citenamefont
  {Schmitt}, \citenamefont {Ficker}, \citenamefont {Wolff}, \citenamefont
  {K\"onig}, \citenamefont {Sizmann},\ and\ \citenamefont
  {Leuchs}}]{Schmidt1998}%
  \BibitemOpen
  \bibfield  {author} {\bibinfo {author} {\bibfnamefont {S.}~\bibnamefont
  {Schmitt}}, \bibinfo {author} {\bibfnamefont {J.}~\bibnamefont {Ficker}},
  \bibinfo {author} {\bibfnamefont {M.}~\bibnamefont {Wolff}}, \bibinfo
  {author} {\bibfnamefont {F.}~\bibnamefont {K\"onig}}, \bibinfo {author}
  {\bibfnamefont {A.}~\bibnamefont {Sizmann}},\ and\ \bibinfo {author}
  {\bibfnamefont {G.}~\bibnamefont {Leuchs}},\ }\bibfield  {title} {\bibinfo
  {title} {Photon-number squeezed solitons from an asymmetric fiber-optic
  sagnac interferometer},\ }\href {https://doi.org/10.1103/PhysRevLett.81.2446}
  {\bibfield  {journal} {\bibinfo  {journal} {Phys. Rev. Lett.}\ }\textbf
  {\bibinfo {volume} {81}},\ \bibinfo {pages} {2446} (\bibinfo {year}
  {1998})}\BibitemShut {NoStop}%
\bibitem [{\citenamefont {Silberhorn}\ \emph {et~al.}(2001)\citenamefont
  {Silberhorn}, \citenamefont {Lam}, \citenamefont {Wei\ss{}}, \citenamefont
  {K\"onig}, \citenamefont {Korolkova},\ and\ \citenamefont
  {Leuchs}}]{Silberhorn2001}%
  \BibitemOpen
  \bibfield  {author} {\bibinfo {author} {\bibfnamefont {C.}~\bibnamefont
  {Silberhorn}}, \bibinfo {author} {\bibfnamefont {P.~K.}\ \bibnamefont {Lam}},
  \bibinfo {author} {\bibfnamefont {O.}~\bibnamefont {Wei\ss{}}}, \bibinfo
  {author} {\bibfnamefont {F.}~\bibnamefont {K\"onig}}, \bibinfo {author}
  {\bibfnamefont {N.}~\bibnamefont {Korolkova}},\ and\ \bibinfo {author}
  {\bibfnamefont {G.}~\bibnamefont {Leuchs}},\ }\bibfield  {title} {\bibinfo
  {title} {Generation of continuous variable einstein-podolsky-rosen
  entanglement via the kerr nonlinearity in an optical fiber},\ }\href
  {https://doi.org/10.1103/PhysRevLett.86.4267} {\bibfield  {journal} {\bibinfo
   {journal} {Phys. Rev. Lett.}\ }\textbf {\bibinfo {volume} {86}},\ \bibinfo
  {pages} {4267} (\bibinfo {year} {2001})}\BibitemShut {NoStop}%
\bibitem [{\citenamefont {Korolkova}\ \emph {et~al.}(2002)\citenamefont
  {Korolkova}, \citenamefont {Leuchs}, \citenamefont {Loudon}, \citenamefont
  {Ralph},\ and\ \citenamefont {Silberhorn}}]{Korolkova2002}%
  \BibitemOpen
  \bibfield  {author} {\bibinfo {author} {\bibfnamefont {N.}~\bibnamefont
  {Korolkova}}, \bibinfo {author} {\bibfnamefont {G.}~\bibnamefont {Leuchs}},
  \bibinfo {author} {\bibfnamefont {R.}~\bibnamefont {Loudon}}, \bibinfo
  {author} {\bibfnamefont {T.~C.}\ \bibnamefont {Ralph}},\ and\ \bibinfo
  {author} {\bibfnamefont {C.}~\bibnamefont {Silberhorn}},\ }\bibfield  {title}
  {\bibinfo {title} {Polarization squeezing and continuous-variable
  polarization entanglement},\ }\href
  {https://doi.org/10.1103/PhysRevA.65.052306} {\bibfield  {journal} {\bibinfo
  {journal} {Phys. Rev. A}\ }\textbf {\bibinfo {volume} {65}},\ \bibinfo
  {pages} {052306} (\bibinfo {year} {2002})}\BibitemShut {NoStop}%
\bibitem [{\citenamefont {Korolkova}\ \emph {et~al.}(2001)\citenamefont
  {Korolkova}, \citenamefont {Loudon}, \citenamefont {Gardavsky}, \citenamefont
  {Hamilton},\ and\ \citenamefont {Leuchs}}]{Korolkova2001}%
  \BibitemOpen
  \bibfield  {author} {\bibinfo {author} {\bibfnamefont {N.}~\bibnamefont
  {Korolkova}}, \bibinfo {author} {\bibfnamefont {R.}~\bibnamefont {Loudon}},
  \bibinfo {author} {\bibfnamefont {G.}~\bibnamefont {Gardavsky}}, \bibinfo
  {author} {\bibfnamefont {M.~W.}\ \bibnamefont {Hamilton}},\ and\ \bibinfo
  {author} {\bibfnamefont {G.}~\bibnamefont {Leuchs}},\ }\bibfield  {title}
  {\bibinfo {title} {Time evolution of a quantum soliton in a kerr medium},\
  }\href {https://doi.org/10.1080/09500340108232466} {\bibfield  {journal}
  {\bibinfo  {journal} {Journal of Modern Optics}\ }\textbf {\bibinfo {volume}
  {48}},\ \bibinfo {pages} {1339} (\bibinfo {year} {2001})},\ \Eprint
  {https://arxiv.org/abs/https://www.tandfonline.com/doi/pdf/10.1080/09500340108232466}
  {https://www.tandfonline.com/doi/pdf/10.1080/09500340108232466} \BibitemShut
  {NoStop}%
\bibitem [{\citenamefont {Haus}\ and\ \citenamefont {Lai}(1990)}]{Haus1990}%
  \BibitemOpen
  \bibfield  {author} {\bibinfo {author} {\bibfnamefont {H.~A.}\ \bibnamefont
  {Haus}}\ and\ \bibinfo {author} {\bibfnamefont {Y.}~\bibnamefont {Lai}},\
  }\bibfield  {title} {\bibinfo {title} {{Quantum theory of soliton squeezing:
  a linearized approach}},\ }\href {https://doi.org/10.1364/JOSAB.7.000386}
  {\bibfield  {journal} {\bibinfo  {journal} {J. Opt. Soc. Am. B}\ }\textbf
  {\bibinfo {volume} {7}},\ \bibinfo {pages} {386} (\bibinfo {year}
  {1990})}\BibitemShut {NoStop}%
\bibitem [{\citenamefont {Imoto}\ \emph {et~al.}(1985)\citenamefont {Imoto},
  \citenamefont {Haus},\ and\ \citenamefont {Yamamoto}}]{Imoto1985}%
  \BibitemOpen
  \bibfield  {author} {\bibinfo {author} {\bibfnamefont {N.}~\bibnamefont
  {Imoto}}, \bibinfo {author} {\bibfnamefont {H.~A.}\ \bibnamefont {Haus}},\
  and\ \bibinfo {author} {\bibfnamefont {Y.}~\bibnamefont {Yamamoto}},\
  }\bibfield  {title} {\bibinfo {title} {Quantum nondemolition measurement of
  the photon number via the optical kerr effect},\ }\href
  {https://doi.org/10.1103/PhysRevA.32.2287} {\bibfield  {journal} {\bibinfo
  {journal} {Phys. Rev. A}\ }\textbf {\bibinfo {volume} {32}},\ \bibinfo
  {pages} {2287} (\bibinfo {year} {1985})}\BibitemShut {NoStop}%
\bibitem [{\citenamefont {Werner}\ and\ \citenamefont
  {Friberg}(1997)}]{Werner1997}%
  \BibitemOpen
  \bibfield  {author} {\bibinfo {author} {\bibfnamefont {M.~J.}\ \bibnamefont
  {Werner}}\ and\ \bibinfo {author} {\bibfnamefont {S.~R.}\ \bibnamefont
  {Friberg}},\ }\bibfield  {title} {\bibinfo {title} {Phase transitions and the
  internal noise structure of nonlinear schr\"odinger equation solitons},\
  }\href {https://doi.org/10.1103/PhysRevLett.79.4143} {\bibfield  {journal}
  {\bibinfo  {journal} {Phys. Rev. Lett.}\ }\textbf {\bibinfo {volume} {79}},\
  \bibinfo {pages} {4143} (\bibinfo {year} {1997})}\BibitemShut {NoStop}%
\bibitem [{\citenamefont {Werner}(1996)}]{Werner1996}%
  \BibitemOpen
  \bibfield  {author} {\bibinfo {author} {\bibfnamefont {M.~J.}\ \bibnamefont
  {Werner}},\ }\bibfield  {title} {\bibinfo {title} {Quantum statistics of
  fundamental and higher-order coherent quantum solitons in raman-active
  waveguides},\ }\href {https://doi.org/10.1103/PhysRevA.54.R2567} {\bibfield
  {journal} {\bibinfo  {journal} {Phys. Rev. A}\ }\textbf {\bibinfo {volume}
  {54}},\ \bibinfo {pages} {R2567} (\bibinfo {year} {1996})}\BibitemShut
  {NoStop}%
\bibitem [{\citenamefont {Adami}\ and\ \citenamefont
  {Steeg}(2014)}]{Adami2014}%
  \BibitemOpen
  \bibfield  {author} {\bibinfo {author} {\bibfnamefont {C.}~\bibnamefont
  {Adami}}\ and\ \bibinfo {author} {\bibfnamefont {G.~V.}\ \bibnamefont
  {Steeg}},\ }\bibfield  {title} {\bibinfo {title} {Classical information
  transmission capacity of quantum black holes},\ }\href
  {https://doi.org/10.1088/0264-9381/31/7/075015} {\bibfield  {journal}
  {\bibinfo  {journal} {Classical and Quantum Gravity}\ }\textbf {\bibinfo
  {volume} {31}},\ \bibinfo {pages} {075015} (\bibinfo {year}
  {2014})}\BibitemShut {NoStop}%
\bibitem [{\citenamefont {Couteau}(2018)}]{Couteau2018}%
  \BibitemOpen
  \bibfield  {author} {\bibinfo {author} {\bibfnamefont {C.}~\bibnamefont
  {Couteau}},\ }\bibfield  {title} {\bibinfo {title} {Spontaneous parametric
  down-conversion},\ }\href {https://doi.org/10.1080/00107514.2018.1488463}
  {\bibfield  {journal} {\bibinfo  {journal} {Contemporary Physics}\ }\textbf
  {\bibinfo {volume} {59}},\ \bibinfo {pages} {291} (\bibinfo {year} {2018})},\
  \Eprint {https://arxiv.org/abs/https://doi.org/10.1080/00107514.2018.1488463}
  {https://doi.org/10.1080/00107514.2018.1488463} \BibitemShut {NoStop}%
\bibitem [{\citenamefont {Agrawal}(2019)}]{Agrawal2019}%
  \BibitemOpen
  \bibfield  {author} {\bibinfo {author} {\bibfnamefont {G.~P.}\ \bibnamefont
  {Agrawal}},\ }\bibfield  {title} {\bibinfo {title} {{Chapter 2 - Pulse
  propagation in fibers}},\ }in\ \href
  {https://doi.org/https://doi.org/10.1016/B978-0-12-817042-7.00009-9} {\emph
  {\bibinfo {booktitle} {Nonlinear Fiber Optics (Sixth Edition)}}},\ \bibinfo
  {editor} {edited by\ \bibinfo {editor} {\bibfnamefont {G.~P.}\ \bibnamefont
  {Agrawal}}}\ (\bibinfo  {publisher} {Academic Press},\ \bibinfo {year}
  {2019})\ \bibinfo {edition} {sixth edition}\ ed.,\ pp.\ \bibinfo {pages}
  {27--55}\BibitemShut {NoStop}%
\bibitem [{\citenamefont {Wright}(1991)}]{Wright1991}%
  \BibitemOpen
  \bibfield  {author} {\bibinfo {author} {\bibfnamefont {E.~M.}\ \bibnamefont
  {Wright}},\ }\bibfield  {title} {\bibinfo {title} {{Quantum theory of soliton
  propagation in an optical fiber using the Hartree approximation}},\ }\href
  {https://doi.org/10.1103/PhysRevA.43.3836} {\bibfield  {journal} {\bibinfo
  {journal} {Phys. Rev. A}\ }\textbf {\bibinfo {volume} {43}},\ \bibinfo
  {pages} {3836} (\bibinfo {year} {1991})}\BibitemShut {NoStop}%
\bibitem [{\citenamefont {Lai}(1993)}]{Lai1993}%
  \BibitemOpen
  \bibfield  {author} {\bibinfo {author} {\bibfnamefont {Y.}~\bibnamefont
  {Lai}},\ }\bibfield  {title} {\bibinfo {title} {{Quantum theory of soliton
  propagation: a unified approach based on the linearization approximation}},\
  }\href {https://doi.org/10.1364/JOSAB.10.000475} {\bibfield  {journal}
  {\bibinfo  {journal} {J. Opt. Soc. Am. B}\ }\textbf {\bibinfo {volume}
  {10}},\ \bibinfo {pages} {475} (\bibinfo {year} {1993})}\BibitemShut
  {NoStop}%
\bibitem [{\citenamefont {Margalit}\ and\ \citenamefont
  {Haus}(1998)}]{Margalit1998}%
  \BibitemOpen
  \bibfield  {author} {\bibinfo {author} {\bibfnamefont {M.}~\bibnamefont
  {Margalit}}\ and\ \bibinfo {author} {\bibfnamefont {H.~A.}\ \bibnamefont
  {Haus}},\ }\bibfield  {title} {\bibinfo {title} {{Accounting for the
  continuum in analysis of squeezing with solitons}},\ }\href
  {https://doi.org/10.1364/JOSAB.15.001387} {\bibfield  {journal} {\bibinfo
  {journal} {J. Opt. Soc. Am. B}\ }\textbf {\bibinfo {volume} {15}},\ \bibinfo
  {pages} {1387} (\bibinfo {year} {1998})}\BibitemShut {NoStop}%
\bibitem [{\citenamefont {Haus}\ and\ \citenamefont {Yu}(2000)}]{Haus2000}%
  \BibitemOpen
  \bibfield  {author} {\bibinfo {author} {\bibfnamefont {H.~A.}\ \bibnamefont
  {Haus}}\ and\ \bibinfo {author} {\bibfnamefont {C.~X.}\ \bibnamefont {Yu}},\
  }\bibfield  {title} {\bibinfo {title} {{Soliton squeezing and the
  continuum}},\ }\href {https://doi.org/10.1364/JOSAB.17.000618} {\bibfield
  {journal} {\bibinfo  {journal} {J. Opt. Soc. Am. B}\ }\textbf {\bibinfo
  {volume} {17}},\ \bibinfo {pages} {618} (\bibinfo {year} {2000})}\BibitemShut
  {NoStop}%
\bibitem [{\citenamefont {Kaup}(1990)}]{Kaup1990}%
  \BibitemOpen
  \bibfield  {author} {\bibinfo {author} {\bibfnamefont {D.~J.}\ \bibnamefont
  {Kaup}},\ }\bibfield  {title} {\bibinfo {title} {{Perturbation theory for
  solitons in optical fibers}},\ }\href
  {https://doi.org/10.1103/PhysRevA.42.5689} {\bibfield  {journal} {\bibinfo
  {journal} {Phys. Rev. A}\ }\textbf {\bibinfo {volume} {42}},\ \bibinfo
  {pages} {5689} (\bibinfo {year} {1990})}\BibitemShut {NoStop}%
\bibitem [{\citenamefont {Kaup}(1991)}]{Kaup1991}%
  \BibitemOpen
  \bibfield  {author} {\bibinfo {author} {\bibfnamefont {D.~J.}\ \bibnamefont
  {Kaup}},\ }\bibfield  {title} {\bibinfo {title} {{Second-order perturbations
  for solitons in optical fibers}},\ }\href
  {https://doi.org/10.1103/PhysRevA.44.4582} {\bibfield  {journal} {\bibinfo
  {journal} {Phys. Rev. A}\ }\textbf {\bibinfo {volume} {44}},\ \bibinfo
  {pages} {4582} (\bibinfo {year} {1991})}\BibitemShut {NoStop}%
\bibitem [{\citenamefont {Sch{\"u}tzhold}\ \emph {et~al.}(2006)\citenamefont
  {Sch{\"u}tzhold}, \citenamefont {Uhlmann}, \citenamefont {Xu},\ and\
  \citenamefont {Fischer}}]{Ralf2006}%
  \BibitemOpen
  \bibfield  {author} {\bibinfo {author} {\bibfnamefont {R.}~\bibnamefont
  {Sch{\"u}tzhold}}, \bibinfo {author} {\bibfnamefont {M.}~\bibnamefont
  {Uhlmann}}, \bibinfo {author} {\bibfnamefont {Y.}~\bibnamefont {Xu}},\ and\
  \bibinfo {author} {\bibfnamefont {U.~R.}\ \bibnamefont {Fischer}},\
  }\bibfield  {title} {\bibinfo {title} {{Mean-field expansion in Bose-Einstein
  condensates with finite-range interactions}},\ }\href
  {https://doi.org/10.1142/S0217979206035631} {\bibfield  {journal} {\bibinfo
  {journal} {International Journal of Modern Physics B}\ }\textbf {\bibinfo
  {volume} {20}},\ \bibinfo {pages} {3555} (\bibinfo {year}
  {2006})}\BibitemShut {NoStop}%
\bibitem [{\citenamefont {Leonhardt}(2010)}]{Leonhardt2010}%
  \BibitemOpen
  \bibfield  {author} {\bibinfo {author} {\bibfnamefont {U.}~\bibnamefont
  {Leonhardt}},\ }\href@noop {} {\emph {\bibinfo {title} {{Essential Quantum
  Optics: From Quantum Measurements to Black Holes}}}}\ (\bibinfo  {publisher}
  {Cambridge University Press},\ \bibinfo {year} {2010})\BibitemShut {NoStop}%
\bibitem [{\citenamefont
  {Mostafazadeh}(2002{\natexlab{a}})}]{Mostafazadeh2002}%
  \BibitemOpen
  \bibfield  {author} {\bibinfo {author} {\bibfnamefont {A.}~\bibnamefont
  {Mostafazadeh}},\ }\bibfield  {title} {\bibinfo {title} {{Pseudo-Hermiticity
  versus PT symmetry: The necessary condition for the reality of the spectrum
  of a non-Hermitian Hamiltonian}},\ }\href {https://doi.org/10.1063/1.1418246}
  {\bibfield  {journal} {\bibinfo  {journal} {Journal of Mathematical Physics}\
  }\textbf {\bibinfo {volume} {43}},\ \bibinfo {pages} {205} (\bibinfo {year}
  {2002}{\natexlab{a}})}\BibitemShut {NoStop}%
\bibitem [{\citenamefont
  {Mostafazadeh}(2002{\natexlab{b}})}]{Mostafazadeh2002II}%
  \BibitemOpen
  \bibfield  {author} {\bibinfo {author} {\bibfnamefont {A.}~\bibnamefont
  {Mostafazadeh}},\ }\bibfield  {title} {\bibinfo {title} {{Pseudo-Hermiticity
  versus PT-symmetry. II. A complete characterization of non-Hermitian
  Hamiltonians with a real spectrum}},\ }\href
  {https://doi.org/10.1063/1.1461427} {\bibfield  {journal} {\bibinfo
  {journal} {Journal of Mathematical Physics}\ }\textbf {\bibinfo {volume}
  {43}},\ \bibinfo {pages} {2814} (\bibinfo {year} {2002}{\natexlab{b}})},\
  \Eprint
  {https://arxiv.org/abs/https://pubs.aip.org/aip/jmp/article-pdf/43/5/2814/19281624/2814\_1\_online.pdf}
  {https://pubs.aip.org/aip/jmp/article-pdf/43/5/2814/19281624/2814\_1\_online.pdf}
  \BibitemShut {NoStop}%
\bibitem [{\citenamefont
  {Mostafazadeh}(2002{\natexlab{c}})}]{Mostafazadeh2002III}%
  \BibitemOpen
  \bibfield  {author} {\bibinfo {author} {\bibfnamefont {A.}~\bibnamefont
  {Mostafazadeh}},\ }\bibfield  {title} {\bibinfo {title} {{Pseudo-Hermiticity
  versus PT-symmetry III: Equivalence of pseudo-Hermiticity and the presence of
  antilinear symmetries}},\ }\href {https://doi.org/10.1063/1.1489072}
  {\bibfield  {journal} {\bibinfo  {journal} {Journal of Mathematical Physics}\
  }\textbf {\bibinfo {volume} {43}},\ \bibinfo {pages} {3944} (\bibinfo {year}
  {2002}{\natexlab{c}})},\ \Eprint
  {https://arxiv.org/abs/https://pubs.aip.org/aip/jmp/article-pdf/43/8/3944/19034234/3944\_1\_online.pdf}
  {https://pubs.aip.org/aip/jmp/article-pdf/43/8/3944/19034234/3944\_1\_online.pdf}
  \BibitemShut {NoStop}%
\bibitem [{\citenamefont {Mostafazadeh}(2004)}]{Mostafazadeh2004}%
  \BibitemOpen
  \bibfield  {author} {\bibinfo {author} {\bibfnamefont {A.}~\bibnamefont
  {Mostafazadeh}},\ }\bibfield  {title} {\bibinfo {title} {{Pseudounitary
  operators and pseudounitary quantum dynamics}},\ }\href
  {https://doi.org/10.1063/1.1646448} {\bibfield  {journal} {\bibinfo
  {journal} {Journal of Mathematical Physics}\ }\textbf {\bibinfo {volume}
  {45}},\ \bibinfo {pages} {932} (\bibinfo {year} {2004})}\BibitemShut
  {NoStop}%
\bibitem [{\citenamefont {Mostafazadeh}(2010)}]{Mostafazadeh2010}%
  \BibitemOpen
  \bibfield  {author} {\bibinfo {author} {\bibfnamefont {A.}~\bibnamefont
  {Mostafazadeh}},\ }\bibfield  {title} {\bibinfo {title} {{Pseudo-Hermitian
  Representation of Quantum Mechanics}},\ }\href
  {https://doi.org/10.1142/S0219887810004816} {\bibfield  {journal} {\bibinfo
  {journal} {International Journal of Geometric Methods in Modern Physics}\
  }\textbf {\bibinfo {volume} {07}},\ \bibinfo {pages} {1191} (\bibinfo {year}
  {2010})}\BibitemShut {NoStop}%
\bibitem [{\citenamefont {Mostafazadeh}(2013)}]{Mostafazadeh2013}%
  \BibitemOpen
  \bibfield  {author} {\bibinfo {author} {\bibfnamefont {A.}~\bibnamefont
  {Mostafazadeh}},\ }\bibfield  {title} {\bibinfo {title} {{Pseudo-Hermitian
  quantum mechanics with unbounded metric operators}},\ }\href
  {https://doi.org/10.1098/rsta.2012.0050} {\bibfield  {journal} {\bibinfo
  {journal} {Philosophical Transactions of the Royal Society A: Mathematical,
  Physical and Engineering Sciences}\ }\textbf {\bibinfo {volume} {371}},\
  \bibinfo {pages} {20120050} (\bibinfo {year} {2013})}\BibitemShut {NoStop}%
\bibitem [{\citenamefont {Bender}\ \emph {et~al.}(2003)\citenamefont {Bender},
  \citenamefont {Brody},\ and\ \citenamefont {Jones}}]{Bender2003}%
  \BibitemOpen
  \bibfield  {author} {\bibinfo {author} {\bibfnamefont {C.~M.}\ \bibnamefont
  {Bender}}, \bibinfo {author} {\bibfnamefont {D.~C.}\ \bibnamefont {Brody}},\
  and\ \bibinfo {author} {\bibfnamefont {H.~F.}\ \bibnamefont {Jones}},\
  }\bibfield  {title} {\bibinfo {title} {{Must a Hamiltonian be Hermitian?}},\
  }\href {https://doi.org/10.1119/1.1574043} {\bibfield  {journal} {\bibinfo
  {journal} {American Journal of Physics}\ }\textbf {\bibinfo {volume} {71}},\
  \bibinfo {pages} {1095} (\bibinfo {year} {2003})}\BibitemShut {NoStop}%
\bibitem [{\citenamefont {Bender}\ and\ \citenamefont
  {Mannheim}(2008)}]{Bender2008}%
  \BibitemOpen
  \bibfield  {author} {\bibinfo {author} {\bibfnamefont {C.~M.}\ \bibnamefont
  {Bender}}\ and\ \bibinfo {author} {\bibfnamefont {P.~D.}\ \bibnamefont
  {Mannheim}},\ }\bibfield  {title} {\bibinfo {title} {{Exactly solvable
  $\mathcal{P}\mathcal{T}$-symmetric Hamiltonian having no Hermitian
  counterpart}},\ }\href {https://doi.org/10.1103/PhysRevD.78.025022}
  {\bibfield  {journal} {\bibinfo  {journal} {Phys. Rev. D}\ }\textbf {\bibinfo
  {volume} {78}},\ \bibinfo {pages} {025022} (\bibinfo {year}
  {2008})}\BibitemShut {NoStop}%
\bibitem [{Note1()}]{Note1}%
  \BibitemOpen
  \bibinfo {note} {In fact, $\protect \pmb {H}_\protect \mathrm
  {BdG}^*=\protect \pmb {H}_\protect \mathrm {BdG}$ here, but we use the
  complex conjugate for to conform to literature~\cite
  {LucaThesis}.}\BibitemShut {Stop}%
\bibitem [{\citenamefont {Yang}(2000)}]{Yang2000}%
  \BibitemOpen
  \bibfield  {author} {\bibinfo {author} {\bibfnamefont {J.}~\bibnamefont
  {Yang}},\ }\bibfield  {title} {\bibinfo {title} {{Complete eigenfunctions of
  linearized integrable equations expanded around a soliton solution}},\ }\href
  {https://doi.org/10.1063/1.1287639} {\bibfield  {journal} {\bibinfo
  {journal} {Journal of Mathematical Physics}\ }\textbf {\bibinfo {volume}
  {41}},\ \bibinfo {pages} {6614} (\bibinfo {year} {2000})},\ \Eprint
  {https://arxiv.org/abs/https://pubs.aip.org/aip/jmp/article-pdf/41/9/6614/11309986/6614\_1\_online.pdf}
  {https://pubs.aip.org/aip/jmp/article-pdf/41/9/6614/11309986/6614\_1\_online.pdf}
  \BibitemShut {NoStop}%
\bibitem [{\citenamefont {Ripka}\ \emph {et~al.}(1986)\citenamefont {Ripka},
  \citenamefont {Blaizot},\ and\ \citenamefont {Ripka}}]{Ripka1986}%
  \BibitemOpen
  \bibfield  {author} {\bibinfo {author} {\bibfnamefont {S.}~\bibnamefont
  {Ripka}}, \bibinfo {author} {\bibfnamefont {J.}~\bibnamefont {Blaizot}},\
  and\ \bibinfo {author} {\bibfnamefont {G.}~\bibnamefont {Ripka}},\
  }\href@noop {} {\emph {\bibinfo {title} {{Quantum Theory of Finite
  Systems}}}}\ (\bibinfo  {publisher} {MIT Press},\ \bibinfo {year}
  {1986})\BibitemShut {NoStop}%
\bibitem [{\citenamefont {Sokolov}\ \emph {et~al.}(2006)\citenamefont
  {Sokolov}, \citenamefont {Andrianov},\ and\ \citenamefont
  {Cannata}}]{Sokolov2006}%
  \BibitemOpen
  \bibfield  {author} {\bibinfo {author} {\bibfnamefont {A.~V.}\ \bibnamefont
  {Sokolov}}, \bibinfo {author} {\bibfnamefont {A.~A.}\ \bibnamefont
  {Andrianov}},\ and\ \bibinfo {author} {\bibfnamefont {F.}~\bibnamefont
  {Cannata}},\ }\bibfield  {title} {\bibinfo {title} {{Non-Hermitian quantum
  mechanics of non-diagonalizable Hamiltonians: puzzles with self-orthogonal
  states}},\ }\href {https://doi.org/10.1088/0305-4470/39/32/S20} {\bibfield
  {journal} {\bibinfo  {journal} {Journal of Physics A: Mathematical and
  General}\ }\textbf {\bibinfo {volume} {39}},\ \bibinfo {pages} {10207}
  (\bibinfo {year} {2006})}\BibitemShut {NoStop}%
\bibitem [{\citenamefont {Giacomelli}(2021)}]{LucaThesis}%
  \BibitemOpen
  \bibfield  {author} {\bibinfo {author} {\bibfnamefont {L.}~\bibnamefont
  {Giacomelli}},\ }\emph {\bibinfo {title} {Superradiant phenomena - Lessons
  from and for Bose-Einstein condensates}},\ \href@noop {} {\bibinfo {type}
  {Phd thesis}},\ \bibinfo  {school} {Universit\`a degli studi di Trento},
  \bibinfo {address} {Trento TN} (\bibinfo {year} {2021}),\ \bibinfo {note}
  {available at \url{https://hdl.handle.net/11572/294551}}\BibitemShut
  {NoStop}%
\end{thebibliography}%
\appendix
\begin{widetext}

\section{Calculation for Classical Field Correction}
\label{App:CorrectionCalculation}
In this appendix, we show how to calculate the particular solution for the classical field correction. The superscript $(p)$ will be omitted for notational simplicity.
By adding and subtracting the first and second row of Eq.~\eqref{eq:correction}, we obtain
\begin{align}\label{eq:realimaginaryeq1}
    \partial_tR-\left(-\frac{1}{2}\partial_x^2-\sech^2\!\!x+\frac{1}{2}\right)I&:=\partial_tR-\hat{D}_1I=S_I(t,x),\\
    -\partial_tI-\left(-\frac{1}{2}\partial_x^2-3\sech^2\!\!x+\frac{1}{2}\right)R&:=-\partial_tI-\hat{D}_3R=-\frac{1}{4}\sech^5x+S_R(t,x)\label{eq:realimaginaryeq2}.
\end{align}
where $\Re[\tilde{E}_\c]=R, \Im[\tilde{E}_\c]=I$.
From Eq.~\eqref{eq:DiscreteRealSource} and \eqref{eq:DiscreteImaginarySource}, we notice that the right hand side of Eq.~\eqref{eq:realimaginaryeq1} and \eqref{eq:realimaginaryeq2} are odd and even polynomial in $t$ each.
Let us set the ansatz
\begin{align}\label{eq:ansatz}
    R(t,x) = \sum_{n=0}^{N/2} f_{2n}(x)t^{2n}, \qquad I(t,x) = \sum_{n=1}^{N/2} g_{2n-1}(x)t^{2n-1}.
\end{align}
where $N\in\{0,2,4,\ldots\}$ is determined by the source and $f_{N}\neq0,\,g_{N-1}\neq0$.
Decoupling Eqs.\eqref{eq:realimaginaryeq1} and \eqref{eq:realimaginaryeq2}, we have
\begin{align}\label{eq:Requation}
    \Big[\partial_t^2+\hat{D}_1\hat{D}_3\Big]R &= -\hat{D}_1S_R+\partial_tS_I+\frac{1}{4}\hat{D}_1\sech^5{x},\\
    \Big[\partial_t^2+\hat{D}_3\hat{D}_1\Big]\,\,I&=\hat{D}_3S_I-\partial_tS_R.\label{eq:Iequation}
\end{align}
From these equations, we will show that the ansatz \eqref{eq:ansatz} is in fact $t$ 4th-order polynomial for our case. 
To show this, let us assume that the highest order of $t$ in the source term is $M<N-1$.
Then $\forall m>M:$
\begin{subequations}\label{eq:ansatzdecoupled}
\begin{align}
    (m+2)(m+1)f_{m+2}+\hat{D}_1\hat{D}_3f_m &=0,\\
    (m+2)(m+1)g_{m+2}+\hat{D}_3\hat{D}_1g_m&=0
\end{align}    
\end{subequations}
And for $N$, we have
\begin{align}
    \hat{D}_1\hat{D}_3f_N &=0,\\
    \hat{D}_3\hat{D}_1g_{N-1}&=0
\end{align}
The possible solution which does not diverge at $x\to\infty$ is
\begin{align}
    f_N &= c_{f1}\tanh{x}\sech{x}+c_{f2} (x\tanh{x}\sech{x}-\sech{x})\\
    g_{N-1}&= c_{g1}\sech{x}+c_{g2} x \sech{x}
\end{align}
If $N>M+3$, from Eq.~\eqref{eq:ansatzdecoupled}, we get
\begin{subequations}\label{eq:ansatzdecoupledHom}
\begin{align}
    N(N-1)f_{N}+\hat{D}_1\hat{D}_3f_{N-2} &=0,\\
    (N-1)(N-2)g_{N-1}+\hat{D}_3\hat{D}_1g_{N-3}&=0
\end{align}    
\end{subequations}
To have the solution which does not diverge at $x\to\infty$, one has the condition $c_{f1}=c_{f2}=c_{g1}=c_{g2}=0$ which yields $f_N=g_{N-1}=0$.
Therefore, $M<N<M+3$.
Recall that our source term from continuous mode contribution is zeroth order in $t$ and discrete mode contribution is quadratic function of $t$.
Therefore, the classical field correction is maximally quadratic and 4th-order of $t$ each.
For our purpose, we solve the continuous and discrete mode contributions separately.
In below, we put superscript $\dis,\con$ to indicate them.

\subsection{Continuous Mode Contribution}
\label{sec:contmode}
The source term coming from the continuous mode contribution satisfies
\begin{align}\label{eq:realimaginaryeq1con}
    \partial_tR_\con-\hat{D}_1I_\con&=0,\\
    -\partial_tI_\con-\hat{D}_3R_\con&=-\frac{1}{4}\sech^5x\label{eq:realimaginaryeq2con}.
\end{align}
Substituting Eq.~\eqref{eq:ansatz} with $N=2$ into Eq.~\eqref{eq:realimaginaryeq1con} and Eq.~\eqref{eq:realimaginaryeq2con}, we get a set of equations: 
\begin{align}
    -\hat{D}_3f_{2,\con}&=0\\
    \hat{D}_1g_{1,\con}&=2f_{2,\con}\\
    -\hat{D}_3f_{0,\con}&=-\frac{1}{4}\sech^5x+g_{1,\con}
\end{align}
The solution is
\begin{subequations}
\label{eq:conparticular}
\begin{align}
    f_{2,\con}(x)&=0\\
    g_{1,\con}(x)&=-\frac{1}{9}  \sech{x}\\
    f_{0,\con}(x)&=c \tanh{x} \sech{x}+\frac{1}{12}  \sech^3\!\!{x}-\frac{5}{18}  \sech{x}+\frac{1}{9}  x \tanh{x} \sech{x},
\end{align}  
\end{subequations}
which is easily verified and where $c$ is an arbitrary real constant.
The term containing c is solves the homogeneous equation and thus, for the purpose of obtaining a particular solution, can be put to zero. [ We can choose $c=0$ and put the contribution into the general solution Eq.~\eqref{eq:correctionfieldExpansion}.
\subsection{Discrete mode contribution}
\label{sec:discrete}
We treat the real and imaginary source terms separately.
For the source term coming from the discrete mode, from Eq.~ \eqref{eq:realimaginaryeq1} and Eq.~\eqref{eq:realimaginaryeq2}, we have 
\begin{align}\label{eq:realimaginaryeqdisc1}
    \partial_tR_\dis-\hat{D}_1I_\dis&=t S_1(x),\\
    -\partial_tI_\dis-\hat{D}_3R_\dis&=S_0(x)+t^2S_2(x).\label{eq:realimaginaryeqdisc2}
\end{align}
where
\begin{align}
    S_2(x)&:=-\frac{1}{4}(1+3\tanh^2\!\!x)\sech^3\!\!{x},\\
    S_1(x)&:=-\frac{1}{2} \, \sech^3x.,\\
    S_0(x)&:=-\frac{1}{4}\big(8+x^2-14x\tanh{x}+3(x^2+1)\tanh^2\!\!x\big)\sech^3\!\!x.
\end{align}
Substituting Eq.~\eqref{eq:ansatz} with $N=4$ into Eq.~\eqref{eq:realimaginaryeqdisc1} and Eq.~\eqref{eq:realimaginaryeqdisc2}, we write
\begin{align}
    4t^3f_{4,\dis}+2tf_{2,\dis}-\hat{D}_1(g_{1,\dis} t+g_{3,\dis} t^3)&=tS_1,\\
    -3g_{3,\dis} t^2-g_{1,\dis}-\hat{D}_3(f_{0,\dis}+f_{2,\dis} t^2+f_{4,\dis} t^4)&=S_2t^2+S_0.
\end{align}
Hence, we obtain the following set of equations:
\begin{align}
    \hat{D}_3f_{4,\dis}&=0\label{eq:AppDiseqf4}\\
    \hat{D}_1g_{3,\dis}&=4f_{4,\dis},\label{eq:AppDiseqg3}\\
    -\hat{D}_3f_{2,\dis}&=3g_{3,\dis}+S_2,\label{eq:AppDiseqf2}\\
    \hat{D}_1g_{1,\dis}&=2f_{2,\dis}-S_1,\label{eq:AppDiseqg1}\\
    -\hat{D}_3f_{0,\dis}&=g_{1,\dis}+S_0.\label{eq:AppDiseqf0}
\end{align}
Let us solve the first equation \eqref{eq:AppDiseqf4}. The solution, which satisfies the boundary condition, is:
\begin{equation}
    f_{4,\dis}(x) = c_1^{(4)} \tanh{x}\sech{x}.
\end{equation}
Next, we solve Eq.~\eqref{eq:AppDiseqg3}:
\begin{equation}
    g_{3,\dis}(x)=c_1^{(4)} (2 x \sech{x}-2 \sinh{x})+c_1^{(3)} \sech{x}+c_2^{(3)} (\sinh{x}+x \sech{x}).
\end{equation}
For a physical solution, the field asymptotically needs to vanish, thus,
\begin{equation}
    -2 \, c_1^{(4)} + c_2^{(3)}=0.
\end{equation}
Now we have
\begin{align}
    f_{4,\dis}(x)&=\frac{1}{2} c_2^{(3)} \tanh{x} \sech{x},\\
    g_{3,\dis}(x)&=(c_1^{(3)}+2c_2^{(3)}x)\sech{x}.
\end{align}

Next we solve \eqref{eq:AppDiseqf2} and get $f_{2,\dis}$ :
\begin{align}
    f_{2,\dis}(x)&=\frac{1}{2}c_1^{(3)} \left(3 \cosh{x}-3 \sech{x}+3 x \tanh{x} \sech{x}\right)\nonumber\\
    &\quad+c_2^{(3)} \left(-3 x^2 \tanh{x} \sech{x}+3 \sinh{x}+6 x
   \sech{x}-9 \tanh{x} \sech{x}\right)\nonumber\\
   &\quad+c_1^{(2)} \tanh{x} \sech{x}+c_2^{(2)} (\cosh{x}-3 \sech{x}+3 x \tanh{x} \sech{x})\nonumber\\
   &\quad-\frac{1}{8}\cosh
   {x}-\frac{1}{4} \sech^3\!\!{x}+\frac{3}{8} \sech{x}-\frac{3}{8} x \tanh{x} \sech{x}.
\end{align}
The asymptotic condition in this case is
\begin{equation}
    c_1^{(3)}=\frac{1}{12}(1-8c_2^{(2)}),\qquad c_2^{(2)}=0,
\end{equation} 
which results in
\begin{align}
    f_{4,\dis}(x)&=0,\\
    g_{3,\dis}(x)&=\frac{1}{12}(1-8c_2^{(2)})\sech{x},\\
    f_{2,\dis}(x)&=c_1^{(2)} \tanh{x} \sech{x}-2 c_2^{(2)}(1- x \tanh{x} )\sech{x}+\frac{1}{4}(\tanh{x}-x)\tanh{x}\sech{x}.
\end{align}
The equation \eqref{eq:AppDiseqg1} gives
\begin{align}
    g_{1,\dis}(x)&=c_1^{(2)} (x \sech{x}-\sinh{x})+2c_2^{(2)} \big((x^2-1) \sech{x}+\cosh{x}\big)+c_1^{(1)} \sech{x}+c_2^{(1)} (\sinh{x}+x\sech{x})\nonumber\\
    &\quad-\frac{1}{4} \big(x^2 - 1\big) \sech{x}-\frac{1}{4} \cosh{x},
\end{align}
leading to the asymptotic condition
\begin{equation}
    -c_1^{(2)} + c_2^{(1)}=0, \qquad -\frac{1}{4} + 2 c_2^{(2)}=0.
\end{equation}
We use these asymptotic condition to determine the constants $c_2^{(2)}$ and $c_1^{(2)}$. Then the solutions we have are of the form
\begin{align}
    g_{3,\dis}(x)&=0,\\
    f_{2,\dis}(x)&=c_2^{(1)} \tanh{x} \sech{x}-\frac{1}{4} \sech^3x,\\
    g_{1,\dis}(x)&=c_1^{(1)} \sech{x}+2 c_2^{(1)} x \sech{x}.
\end{align}
The final equation \eqref{eq:AppDiseqf0} gives
\begin{align}
    f_{0,\dis}(x)&=\frac{1}{2}c_1^{(1)} \left(\cosh{x}-\sech{x}+x \tanh{x} \sech{x}\right)\nonumber\\
    &\quad+c_2^{(1)} \big(\sinh{x}-(x^2+3) \tanh{x}\sech{x}+2 x
   \sech{x}\big)\nonumber\\
    &\quad+c_1^{(0)} \tanh{x} \sech{x}+c_2^{(0)} (\cosh{x}-3 \sech{x}+3 x \tanh{x} \sech{x})\nonumber\\
    &\quad -\frac{7}{15}\cosh{x}+\frac{4}{15}\big(x \sinh{x}+\cosh{x}\ln{(\sech{x})}\big)-\frac{1}{4} (1+x^2)\sech^3\!\!{x}\nonumber\\
    &\quad -\frac{1}{15} x \big(12 x+12 \ln(\sech{x})+31-24 \ln{2}\big)\tanh{x}\sech{x}-\frac{4}{5} \text{Li}_2\left(-e^{-2 x}\right)\tanh{x}\sech{x}.
\end{align}
The asymptotic condition results in
\begin{equation}
    -\frac{7}{15}+\frac{1}{2}c_1^{(1)}+c_2^{(0)}-\frac{4}{15}\ln{2}=0,\qquad c_2^{(1)}=0.
\end{equation}
Therefore
\begin{align}
    f_{2,\dis}(x)&=-\frac{1}{4} \sech^3x,\\
    g_{1,\dis}(x)&=-2 c_2^{(0)} \sech{x}-\frac{2}{15}(7+4\ln{2})\sech{x},\\
    f_{0,\dis}(x)&=c_1^{(0)} \tanh{x} \sech{x}+2c_2^{(0)} (x \tanh{x} -1) \sech{x}+\frac{4}{15}\Big(x \sinh{x}+\cosh{x}\big(\ln{(\sech{x})}-\ln2\big)\Big)\nonumber\\
    &\quad +\frac{1}{60}\big(15+16\ln2-48\ln (\sech{x})\big)\sech{x}-\frac{1}{4} (1+x^2)\sech^3\!\!{x}\nonumber\\
   &\quad -\frac{4}{15} \Big(3 \text{Li}_2\big(-e^{-2 x}\big)+x \big(3 x+3 \ln (\sech{x})+6-5 \ln{2}\big)\Big)
\end{align}
For simplicity, let us choose $c_1^{(0)}=0,\,c_2^{(0)}=\frac{23}{60}+\frac{16}{15}\ln{2}$. Then, in summary, only $f_{0,\dis}$ and $f_{2,\dis}$ remain:
\begin{subequations}\label{eq:disparticular}
\begin{align}
    f_{4,\dis}(x)&=0,\\
    g_{3,\dis}(x)&=0,\\
    f_{2,\dis}(x)&=-\frac{1}{4} \sech^3x,\\
    g_{1,\dis}(x)&=0,\\
    f_{0,\dis}(x)&=\frac{4}{15}\Big(x \sinh{x}+\cosh{x}\big(\ln{(\sech{x})}-\ln2\big)\Big)\nonumber\\
    &\quad +\frac{1}{60}\big(-41+48\ln2-48\ln (\sech{x})\big)\sech{x}-\frac{1}{4} (1+x^2)\sech^3\!\!{x}\nonumber\\
   &\quad -\frac{2}{15} \Big(6 \text{Li}_2\big(-e^{-2 x}\big)+x \big(6 x+6 \ln (\sech{x})+5-6 \ln{2}\big)\Big)
\end{align}
\end{subequations}
Note that only the real part of the solution is nonzero.
The real and imaginary part of the solution contributes to the photon number density and current each.
Therefore, there is no current contribution from this particular solution.
Moreover, the density will grow quadratically with $t$ because the highest nonzero contribution for the solution is $f_{2,\dis}(x)$.
\end{widetext}
\end{document}